\documentclass[10pt]{iopart}
\pdfoutput=1

\expandafter\let\csname equation*\endcsname\relax

\expandafter\let\csname endequation*\endcsname\relax

\usepackage{amsmath}

\usepackage[utf8]{inputenc}
\usepackage{multirow}

\usepackage[dvips]{graphics,graphicx}
\usepackage{url}
\usepackage{subfigure}
\usepackage{etoolbox}
\usepackage{booktabs}
\usepackage[table,xcdraw]{xcolor}
\usepackage{ulem}
\usepackage{overpic}
\usepackage{hyperref}
\usepackage{amssymb}
\usepackage{color}
\usepackage{tablefootnote}

\definecolor{bananayellow}{rgb}{1.0, 0.88, 0.21}
\definecolor{amethyst}{rgb}{0.6, 0.4, 0.8}
\definecolor{blizzardblue}{rgb}{0.67, 0.9, 0.93}
\definecolor{bluegray}{rgb}{0.4, 0.6, 0.8}
\definecolor{harvestgold}{rgb}{0.85, 0.57, 0.0}
\definecolor{junebud}{rgb}{0.74, 0.85, 0.34}

\newtoggle{draft}

\toggletrue{draft}

\newcommand{\qprof}{\mathsf{q}}

\begin{document}
\title[Efficient and robust evaluation of fast particle losses]{Efficient and robust evaluation of fast particle losses in non-axisymmetric tokamak plasmas}

\author{Konsta S\"arkim\"aki\textsuperscript{1}}

\address{\textsuperscript{1}Aalto University, Espoo, Finland}

\ead{konsta.sarkimaki@aalto.fi}
\vspace{10pt}
\begin{indented}
\item[]\today
\end{indented}

\begin{abstract}\\
We present various techniques that make orbit-following Monte Carlo simulations faster and more reliable when assessing collisionless fast particle losses due to magnetic field perturbations.
These techniques are based on identifying various loss channels in constants of motion space using the so-called loss maps.
We demonstrate that this allows one to attribute losses quantitatively to different transport mechanisms, increase signal-to-noise ratio when estimating FILD signal and peak power loads, and connect magnetic field structure directly to fast particle losses.
Furthermore, we show that collisionless fast particle transport can be treated as an advection-diffusion process where the transport coefficients can be evaluated with the orbit-following method.
Applying these techniques has the potential to make orbit-following simulations faster to perform, or to avoid them completely, while making the results more reliable as they become more clearly connected to underlying physics.
We demonstrate these techniques  for ITER by showing how alpha particle losses are affected by various magnetic field perturbations, estimating ICRH losses in reduced field scenarios, and performing a scan on alpha particle losses as a function of ELM control coil current phases.
\end{abstract}

%
\vspace{2pc}

%

%
%
\ioptwocol

\section{Introduction}
\label{sec:Introduction}

Orbit-following Monte Carlo method is a popular approach to estimate fast particle transport, losses and resulting wall loads in fusion experiments.
The method is well suited for this purpose as it can accurately solve marker trajectories under the influence of 3D magnetic fields.
The method does have two drawbacks: speed and complexity.

Orbit-following calculations can be time-consuming to perform because the accuracy depends on the number of markers to be simulated.
Million markers or more might be required to, e.g., estimate peak power loads\cite{varje2016}.
Therefore, modern orbit-following codes tend to be highly parallelized to the extent that there are codes with multi-level parallelization\cite{varjeascot5} or with full GPU support\cite{akers2012gpgpu}.
This is not always enough to reduce simulation time to acceptable leves, which is why there is a need for additional measures, such as marker splitting\cite{akers2018high} or acceleration of interaction time-scales/time-scale enhancement [ASCOT simulations of fast ion power loads to the plasma-facing components in ITER,Effects of ELM mitigation coils on energetic particle confinement in ITER steady-state]\cite{Kurki_Suonio_2009,Tani_2011}.
However, the marker splitting may yield systematic error while time-scale acceleration is, strickly speaking, valid only in axisymmetric magnetic field.

By calling orbit-following calculations complex, we do not refer to the calculation event itself but how the results are interpreted.
Orbit-following tools are, in the absent of artifical transport coefficient, first-principle codes.
Therefore, one cannot tune out, e.g., ripple-trapping, to see what effect it has on the losses, without modifying the inputs which might cause unintended side-effects.
This is a major issue for the modeller because, while one is always quaranteed to get results, they are not always clear to interpret, or worse, could be due to faulty inputs or code.
It is up for the modeller to find a way to connect results to the underlying physics.

In this contribution, we present novel techniques to address both of these drawbacks and show how the techniques are used in practice.
These techniques are based on the \emph{loss-map analysis} which we originally introduced in Ref.~\cite{sarkimaki2018mechanics}.
The loss-map analysis consists of parametrizing particle orbits according to a specific set of constants of motion.
The analysis is valid when the transport is due to collisionless processes, and the time-scale is short compared to the collision time.
In practice, these limits cover fast ion and runaway electron transport in perturbed magnetic field.
Here we demonstrate these new techniques for ITER fast ion confinement, but they are applicable for other machines as well.

This paper is organized as follows.
The loss-map analysis is outlined in section~\ref{sec:loss map}.
Section~\ref{sec:initialization} describes how to perform marker initialization in support of the loss-map analysis-
In section~\ref{sec:applying loss maps}, we demonstrate the use of loss maps for alpha particles in ITER baseline scenario.
We show how various loss channels appear or are supressed when magnetic field complexity is increased gradually by introducing toroidal field ripple, ferritic inserts, test blanket modules, ELM control coils, and plasma response.
In addition to this, we also explore ICRH generated fast ion confinement in the reduced field scenarios.

Using loss maps offers several prospects which are reviewed in section~\ref{sec:prospects}.
These include: i) showing how one can use magnetic field structure to estimate fast particle losses without carrying out an orbit-following simulation, ii) showing how transport mechanisms, particle's birth position, and location of the power loads are connected iii) improving marker initialization to achieve better signal-to-noise ratio in orbit-following simulations with less markers and shorter simulations times, and iv) identifying what role collisions play in fast ion transport.

Finally, in section~\ref{sec:advection diffusion}, we show how collisionless fast ion transport can be treated as an one-dimensional advection-diffusion process to a good accuracy.
This finding has one immediate application: transport coefficients can be evaluated with an orbit-following code which then can be supplemented to a transport model to estimate fast ion losses.
Here the benefit is that the evaluation of the transport coefficients, and subsequent losses with the transport model, is one or two orders of magnitudes faster than a full slowing-down simulation.
The advection-diffusion model approach is demonstrated by using it to generate a loss map for alpha particles in a realistic ITER magnetic field, and comparing the loss map with the one created with a full orbit-following simulation.
We show that the advection-diffusion model is suitable for performing fast parameter scans for estimating alpha particle losses as a function of ELM control coil phases.

All orbit-following simulations in this contribution are carried out using ASCOT5 code\cite{varjeascot5}.
The code and the post-processing tools are available on request as is the input data.

\section{Loss-map analysis}
\label{sec:loss map}

\begin{figure}[t]
\centering
\begin{overpic}[width=0.3\textwidth]{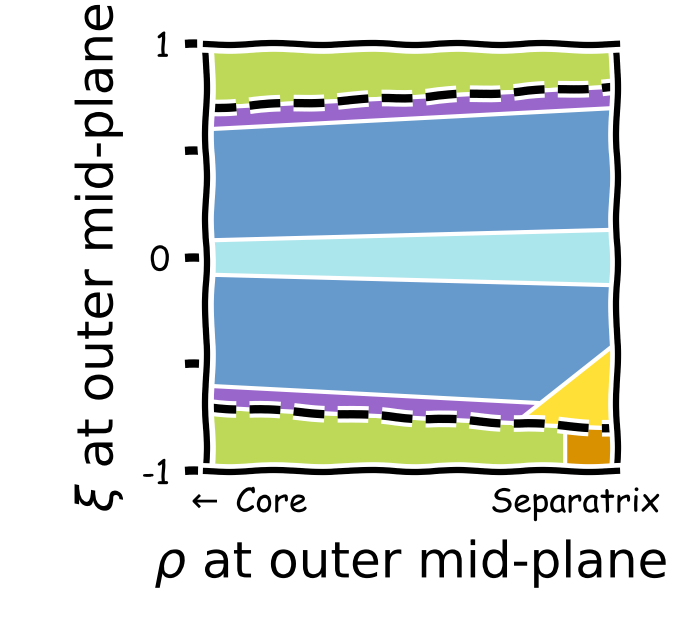}
\end{overpic}
\caption{
Loss channels sketched on $(\rho,\xi)$-plane.
Each colored region corresponds to a different loss or transport mechanism: 
(\textcolor{harvestgold}{orange}) open-field-line losses, 
(\textcolor{bananayellow}{yellow}) gradient-drift losses, 
(\textcolor{bluegray}{blue}) stochastic ripple transport, 
(\textcolor{blizzardblue}{light blue}) ripple-trapping, 
(\textcolor{junebud}{green}) stochastic field-line transport, 
(\textcolor{amethyst}{purple}) perturbed banana transport.
Passing-trapped boundaries are shown with dashed black lines.
}
\label{fig:loss channels illustrated}
\end{figure}

Loss maps are constructed by expressing particle phase-space in a suitable way so that the \emph{loss-channels} can be identified.
Loss channels are regions where all or most particles that are born there are lost.
We show that these regions can be connected to known transport mechanisms, and the purpose of the loss-map analysis is to use this connection in interpreting the results of a orbit-following simulation.
But first, we must find a suitable set of phase-space coordinates to construct loss maps.

Assuming low-collisionality and axisymmetric plasma, particle orbits are determined by three constants of motion. 
These are energy, magnetic moment, and toroidal canonical angular momentum:
\begin{align}
E = \frac{1}{2}mv^2,\\
\mu = \frac{(1-\xi^2)v^2}{B},\\
P_\phi = mRv_\phi + q\psi,
\end{align}
where $m$ is particle mass, $q$ charge, $v$ velocity, $\xi=v_\parallel/v$ pitch, $R$ radial (major radius) coordinate, $B$ magnetic field strength, and $\psi$ poloidal flux.
Fast particles are extremely collisionless with $\omega_b \gg \nu$, where $\omega_b$ is the bounce frequency and $\nu$ the pitch collision frequency, and the constants of motion are conserved for several orbits.
For this reason, neoclassical transport is usually negligible for fast particles.
In non-axisymmetric plasmas, $P_\phi$ is not conserved which opens the door for collisionless transport processes, which can lead to rapid losses of fast particles.

Even in non-axisymmetric plasma, $P_\phi$ can be used to categorize particles with respect to their orbit topology.
This categorization is useful because it is the orbit topology that determines the collisionless transport process a particle is affected by.
Previous work has shown how different loss channels can be identified in $(P,\mu;E)$-space\cite{White_1996,Hsu_1992}.
However, we find a different choice of coordinates to be more intuitive.

We assume that the particles have initially same energy, so that the energy can be treated as a separate parameter.
This assumption is not excessively strict as, e.g., NBI ions are born with distinct energies and fusion-born alpha energies are peaked at 3.5 MeV.
With particle $P$, $\mu$, and $\xi$ given, we can uniquely determine what are the $\psi$ and $\xi$ values \emph{at the location where particle crosses outer mid-plane} (OMP).
These OMP values are denoted with $(\psi',\xi')$.
Trapped particles have two locations where OMP is crossed, but the one where $\xi'$ has the same sign as $\xi$ is used.
To make the choice of coordinates even more convenient, $\psi'$ is replaced with a corresponding normalized coordinate,
\begin{equation}
\rho \equiv \sqrt{\frac{\psi - \psi_\mathrm{axis}}{\psi_\mathrm{sep}-\psi_\mathrm{axis}}}
\end{equation}
where $\psi_\mathrm{axis}$ and $\psi_\mathrm{sep}$ are poloidal flux at the axis and at the separatrix, respectively.
Note that $\rho=1$ at the separatrix and $\rho=0$ at the magnetic axis.

We have chosen $(\rho',\xi')$ to present the particle phase space because the space finite, $(\rho',\xi')\in [0,1]\times[-1,1]$, and because passing-trapped boundary and plasma separatrix are easy to identify.
Figure~\ref{fig:loss channels illustrated} illustrates how loss channels appear in $(\rho',\xi')$ plane.
For reference, we briefly review the different loss channels here:
\begin{itemize}
\item The \emph{first-orbit losses} (orange and yellow), are due to particles that are born with orbits that cross the material surface within first few orbits.
They are present even in axisymmetric plasma.
The \emph{open-field-line losses} are particles who are either born in open field line region, or taken there by the $\nabla B$-drift.
The \emph{gradient-drift losses} are particles whose orbit width is large enough that they become lost.
We make the distinction between these two mechanisms because open-field-line mechanism mostly affect passing particles which are then lost to the divertor, whereas gradient-drift mechanism mostly affect trapped particles which are then lost to the wall.
First-orbit loss channel in Fig.~\ref{fig:loss channels illustrated} is drawn at negative pitch side because, in ITER, those are born in trajectories that open outwards.

\item \emph{Stochastic-ripple transport} arise from toroidal variation of the toroidal field strength\footnote{Here \emph{ripple} refers to any toroidal variation of the toroidal field, whereas \emph{TF ripple} refers specifically to ripple caused by finite number of toroidal field coils.}, which leads to displacement of banana tip points on subsequent reflections\cite{Goldston_1981}.
Assuming the ripple is periodic, the stochastic-ripple transport takes place if the ripple magnitude, $\delta$, exceeds the critical value $\delta > \delta_\mathrm{crit}$.
The ripple magnitude is calculated as,
\begin{equation}
\label{eq:ripple definition}
\delta = \frac{B_\mathrm{max}-B_\mathrm{min}}{B_\mathrm{max}+B_\mathrm{min}},
\end{equation}
where $B_\mathrm{max}$ and $B_\mathrm{min}$ are the toroidal extrema of the toroidal field strength, while the critical value is given by
\begin{equation}
\label{eq:critical ripple}
\delta_\mathrm{crit} = \frac{1}{\rho_g (\partial \qprof/\partial\psi)}\left(\frac{\epsilon}{\pi N \qprof}\right)^{\frac{3}{2}},
\end{equation}
where $N$ is the toroidal mode of the perturbation, $\epsilon = r/R_0$ the inverse aspect ratio, $\rho_g$ the particle gyroradius, and $\qprof$ the safety factor.
The transport is diffusive with diffusion coefficient
\begin{equation}
\label{eq:stochastic ripple diffusion}
D \sim \frac{N\pi\qprof^3 \delta^2\rho_g^2}{\epsilon^3\sin\theta_t}\omega_b,
\end{equation}
where $\theta_t$ is the poloidal angle of the particle banana tip, and $\omega_b=\frac{v_{\perp}}{\qprof R_0}\sqrt{\frac{\epsilon}{2}}$ is the bounce frequency.

\item \emph{Ripple-trapping} consists of particles that are caught in a ripple well and become toroidally trapped.
Since these particles cannot complete their poloidal orbits, they are promptly lost due to the $\nabla B$-drift.
The condition for ripple-trapping is $a^*>1$, where the ripple-well parameter $a^*$ is defined as\cite{Tobita_1992}
\begin{equation}
\label{eq:ripple-well parameter}
a^* \equiv \left|\frac{\frac{\partial \bar{B}}{\partial l}}{\frac{\partial \tilde{B}}{\partial l}}\right|,
\end{equation}
where $\bar{B}$ is the axisymmetric field, $\tilde{B}$ is the non-axisymmetric perturbation, and $\frac{\partial}{\partial l}$ refers to derivate along the field line.
Particles may drift away from the ripple-well region before being lost, and it is difficult to exactly separate this process from the stochastic ripple transport.
Thus, together ripple-trapping and stochastic-ripple transport are referred to as \emph{ripple-induced transport}.

\item \emph{Stochastic-field-line transport} arise when overlapping magnetic islands are formed and the magnetic field becomes ergodic where the flux surfaces used to be.
Passing particles follow these chaotic field lines and the particle radial motion becomes diffusive with coefficient\cite{Rechester_1978}
\begin{equation}
\label{eq:stochastic field diffusion}
D \sim \omega_b^{-1}v_\parallel^2\tilde{b}^2,
\end{equation}
where $\tilde{b}^2 = \left<(\delta B/B)^2\right>$ is the normalized perturbation variance.

\item \emph{Perturbed banana transport} was identified in simulations performed in Ref.~\cite{sarkimaki2018mechanics}.
This transport arise when there is a toroidal variation in poloidal field strength near the X-point.
The poloidal field near X-point affects orbit widths of marginally trapped particles that cross that region:
it determines the time a particle spends near the X-point, and all this time the $\nabla B$-drift pulls particles orbit wider.
When there is a toroidal variation in poloidal field near the X-point, orbit widths of marginally trapped particles will depend on the toroidal angle at which they cross the X-point region. 
This variation in orbit widths leads to stochastization of turning points and transport, akin to stochastic-ripple transport.
\end{itemize}

\section{Optimized marker initialization}
\label{sec:initialization}

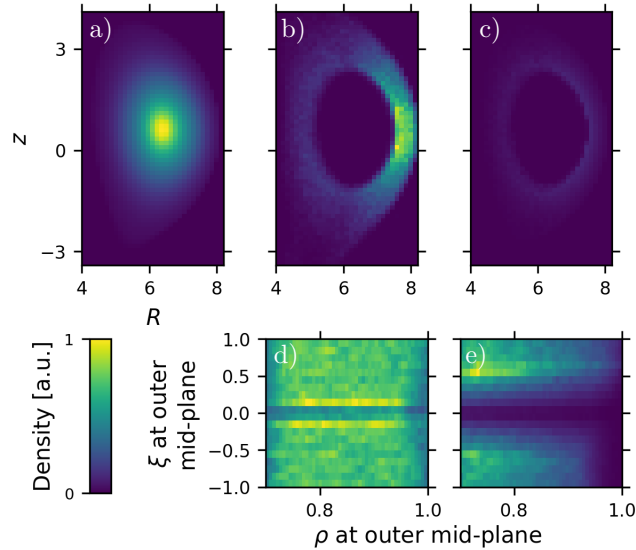
\begin{figure*}[!t]
\centering
\begin{overpic}[width=0.55\textwidth]{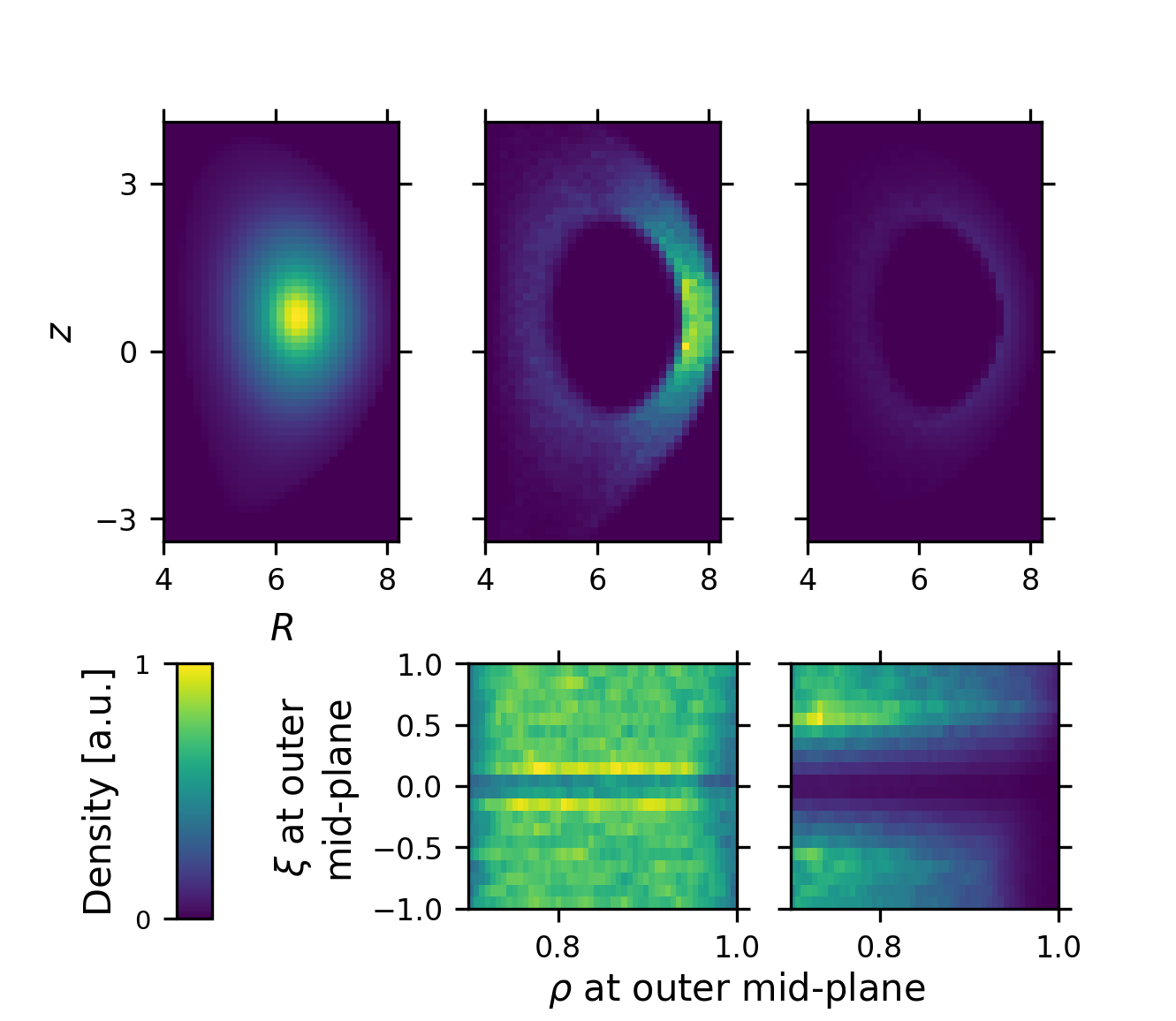}
\put(15,73){\textcolor{white}{a)}}
\put(42,73){\textcolor{white}{b)}}
\put(70,73){\textcolor{white}{c)}}
\put(41,27){\textcolor{white}{d)}}
\put(68,27){\textcolor{white}{e)}}
\end{overpic}
\caption{
Distributions related to marker initialization, optimized for loss calculations, for fusion alphas.
(a) Physical particle distribution which, in this case, is the alpha particle birth rate. 
(b) Marker  distribution chosen so that it is uniform in $(\rho',\xi')$-space and only represents the edge population. 
(c) Weighted marker distribution.
All these distributions are given in $(R,z,\xi)$ but here only the $Rz$-profile is shown.
(d) Marker and (e) weighted marker distribution in $(\rho',\xi')$.
Since weighted marker distribution represents the test particle population at the edge, (e) is same for the population in (a) if only particles at $\rho'>0.8$ are considered.
}
\label{fig:initialization}
\end{figure*}

Before presenting the marker initialization procedure, few terms need to be clarified first because there is some variation in how different authors use them.
\emph{Particles} refer always to physical particles whereas \emph{markers} are pseudo-particles whose orbits are solved with an orbit-following code.
Some authors prefer the term test particle when referring to markers, but here \emph{test particles} refer to the particle population that is to be studied, e.g., fusion alphas.
Each marker represents multiple particles, or particle birth rate in steady-state simulations, and this number is assigned to a marker as a \emph{weight}.
With weights, the marker population represents the test particle population.

We are free to choose our markers as long as the test particle population is accurately represented.
Markers equal CPU time so, preferably, they are chosen as to achieve results with acceptable levels of noise using as few markers as possible.
If losses due to externally induced perturbations are of interest, markers are generated at the edge only.
However, to identify loss channels, the preferred way is to initialize markers so that they are uniformly distributed in $(\rho',\xi')$.

\emph{Marker distribution} is a probability distribution from which marker coordinates are drawn when markers are initialized for the simulation.
In other words, it is how markers are distributed when marker weights are not considered.
One can conceive a marker distribution uniform in $(\rho',\, \xi')$-space by starting from a $(\rho',\, \xi')$-grid, where each node has equal weight (probability).
The nodes are mapped to $(P_\phi,\, \mu)$-space where they appear as a set of points.
Each point in $(P_\phi,\, \mu)$-space still has an equal weight and the next step is to map those points to real-space, $(R,z,\xi)$, which is also divided into a regular grid.
Since each $(P_\phi,\, \mu)$ point corresponds to multiple $(R,z,\xi)$-bins, to all those that lie along the orbit, the weight from a single point is divided evenly to all corresponding bins.
By normalizing the result, we have arrived at a marker distribution in $(R,z,\xi)$-space that is almost (since the transformation is not exact) uniform in $(\rho',\, \xi')$-space.

Figure~\ref{fig:initialization} illustrates how marker distribution is used to obtain weighted markers.
\emph{Particle distribution}, whose $Rz$ profile is shown in (a), represents birth rate of fusion alphas.
Marker distribution in (b) was constructed so that it is uniform in $(\rho',\, \xi')$-space as shown in (d).
Markers are only initialized at $\rho'>0.8$ because earlier simulations have shown that very little losses originate further inside the plasma.
Note that the marker distribution has largest intensity near the OMP:
While all $(\rho',\, \xi')$ values contribute there, the ones with small $\xi'$ deposit there their entire contribution because they correspond to deeply trapped particles.
A requested number of markers is drawn from the marker distribution.
Each marker in a $(R,z,\xi)$ bin is given the correspoinding weight from the particle distribution, divided by the number of markers within that bin.
Resulting \emph{weighted marker distribution} is shown in (c), and its $(\rho',\, \xi')$-space distribution in (e).
Note that because the weighted particle distribution represents the test particle population, distribution in (e) is also how the particle distribution (a) is distributed in $(\rho',\, \xi')$-space.

\section{Applying loss-map analysis to ITER}
\label{sec:applying loss maps}

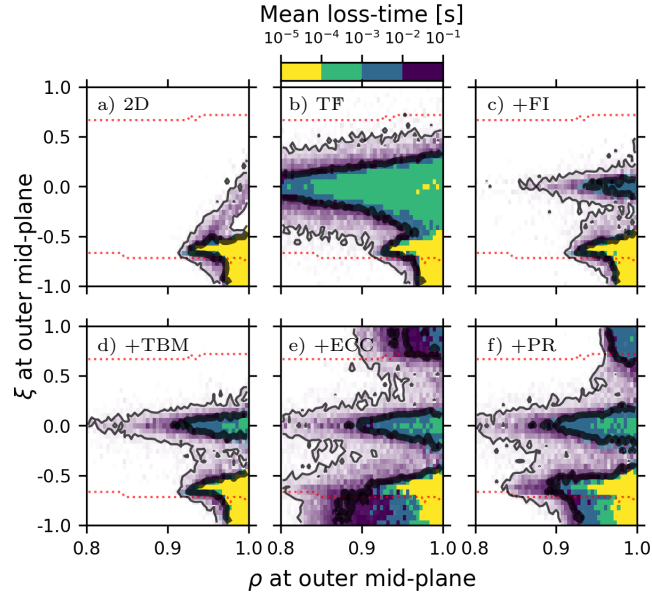
\begin{figure*}[t]
\centering
\begin{overpic}[width=0.55\textwidth]{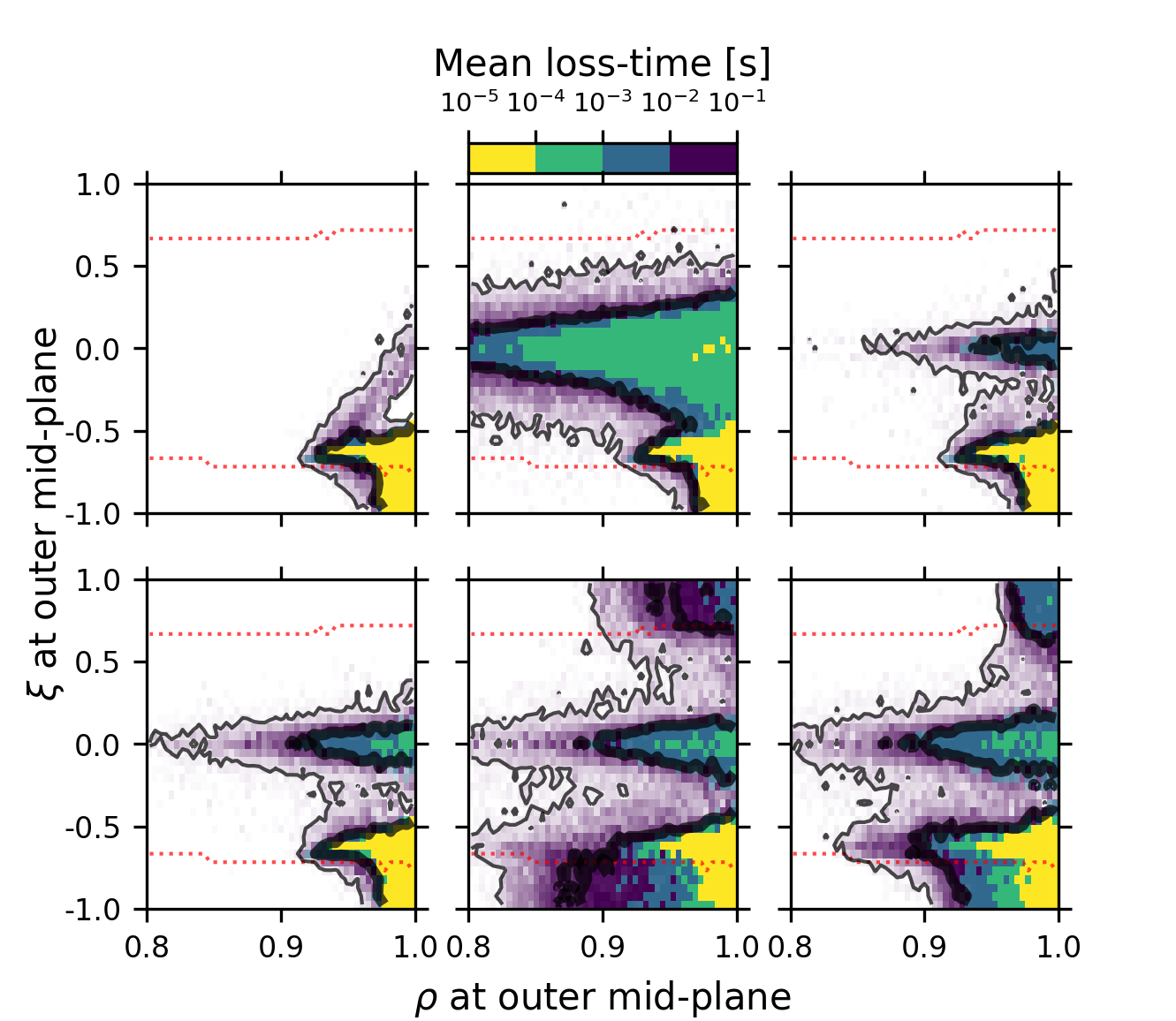}
\put(14,68){\scriptsize{a) 2D}}
\put(41,68){\scriptsize{b) TF}}
\put(69,68){\scriptsize{c) +FI}}
\put(14,34){\scriptsize{d) +TBM}}
\put(41,34){\scriptsize{e) +ECC}}
\put(69,34){\scriptsize{f) +PR}}
\end{overpic}
\caption{
Loss maps for alpha particles during slowing-down in different magnetic configurations.
The plot shows particle birth position in $(\rho',\xi')$ space, and the fraction of markers lost from that region and their mean loss time.
The color shows the mean loss time and the color lightness is varied according to what is the fraction of particles lost locally; no losses occur on regions that appear white.
The black contours provide guidance by showing where over 90~\% (thick line) and 10~\% (thin line) of particles are lost.
The red dotted lines show trapped-passing boundary.
}
\label{fig:loss cases}
\end{figure*}

To demonstrate how constructing loss maps aid in interpreting the results of orbit-following simulations, we carry out alpha particle slowing-down simulations for the ITER baseline scenario.
The purpose of these simulations is to assess how alpha particle losses are connected to various ITER components that cause magnetic field perturbations.
In these simulations, markers are simulated as guiding centers with collisions included.
A simulation lasts until a marker i) hits vessel wall ii) slow-down to 10 keV, or iii) exceeds the simulation time of 0.1 s, during which most losses occur.

The simulations are carried out with various magnetic field perturbations present.
The magnetic field complexity is gradually increased starting from an axisymmetric field (referred as 2D), and then introducing TF-coil ripple (TF), ferritic inserts (+FI), test blanket modules (+TBM), ELM control coils (+ECC) and plasma response (+PR).
These cases were chosen because they represent realistic magnetic field configurations that cover all possible loss-channels.
Furthermore, since fast ion confinement has already been studied in these cases\cite{varje2016,akaslompolo2015iter,kurkisuonio2016protecting,kurkisuonio2016effect}, we can focus on demonstrating the use of loss maps.
For reference, equilibrium and plasma profiles were evaluated in Ref.~\cite{Parail_2013}, perturbation due to TF-coils, FIs and TBMs in Ref.~\cite{akaslompolo2015calculating}, perturbation due to ELM control coils in Ref.~\cite{varje2016}, and plasma response in Ref.~\cite{liu2016modelling}.
The ECC current configuration used here is same as in the previous studies, i.e., $N=3$, $I=45$ kAt, and upper, equatorial, and lower coil phases are $[86^\circ\,0^\circ\,6^\circ]$, which is the configuration that is expected to effectively mitigate ELMs\cite{Evans_2013}.
Visualization of magnetic field structure for these cases can be found in~\ref{app:magnetic}.

Loss-map analysis was already performed for some of the cases in Ref.~\cite{sarkimaki2018mechanics}, but here we make a more thorough presentation.
After performing the orbit-following calculations, markers are divided into $(\rho',\xi')$ bins according to their initial location.
On each bin we calculate the fraction of particles that were lost there during the slowing-down.
As a result, we obtain the loss-maps in Fig.~\ref{fig:loss cases}, which we now proceed to analyze.

Recalling Fig.~\ref{fig:loss channels illustrated}, different loss channels can be identified from the results.
The axisymmetric case is supposed to have only first-orbit losses and collisional losses.
Indeed, (a) has a yellow, finger-like area that corresponds to first orbit losses.
This channel remains unchanged in all cases.
The collisional losses appear as (light) purple because they happen slowly and diffusively.
Collisional losses appear next to the collisionless loss channels in all cases because the collisions scatter particles to these existing loss channels.
Most particles are thermalized before they are scattered to a loss channel, which is why the fraction of particles lost decrease rapidly as one moves away from the loss channels.

Ripple-induced losses (large green region) appear when the TF coil ripple is introduced in (b).
One cannot exactly separate the ripple-trapping and stochastic-ripple transport using just this figure. 
However, one can argue that most of this loss channel is caused by the stochastic-ripple diffusion since the loss-time is comparatively long considering that the ripple-trapping mechanism is advective.
The yellow dots at the very edge, however, could be due to the ripple-trapping.

Introducing FIs (c) and TBMs (d) only affects the ripple-induced loss channel.
FIs are effective at mitigating the ripple and the green region diminishes when those are included.
TBMs cause the green region to grow slightly but not to the extent it was without FIs.
The ripple-induced loss channel is not significantly modified further in +ECC and +PR cases.

So far stochastic-field-line losses have been absent, but they appear once ELM control coils are activated in (e).
Blue and purple regions appear in the passing particle region, and they extent more inwards on the side of the negative pitch (outward-opening orbits).
The time-scale of these losses indicates a diffusive process.
There is also a increase in collisional losses in the trapped particle region.
Introduction of plasma response in (f) suppresses stochastic-field-line losses.
On the other hand, perturbed-banana transport appears which increases transport of marginally trapped particles at $\xi'\approx -0.6$.

\begin{figure*}[t]
\centering
\begin{overpic}[width=0.55\textwidth]{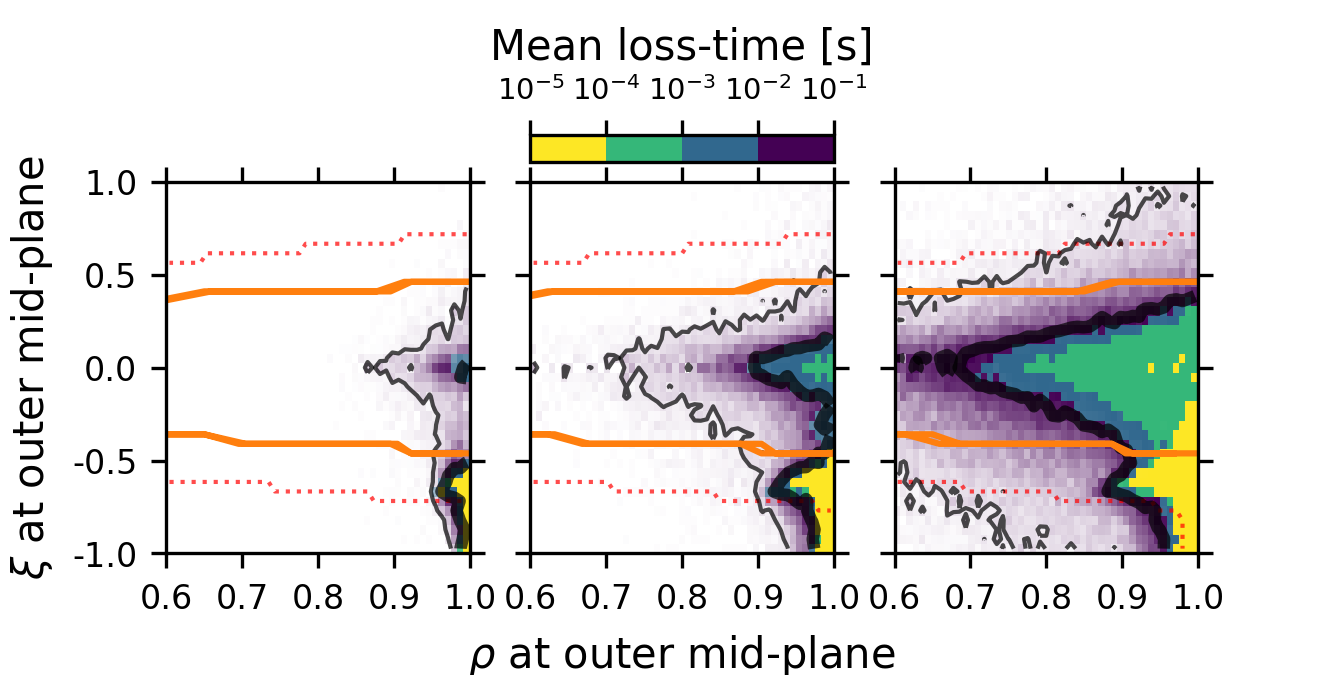}
\put(14,34){a)}
\put(41,34){b)}
\put(69,34){c)}
\end{overpic}
\caption{
Loss maps for 1 MeV  hydrogen ions, representing ICRH ions, in (a) full-field, (b) half-field, and (c) third-field scenarios.
Orange lines marks the location of particles whose banana tip is at $R=R_0$ where the resonance surface is assumed to be.
The meaning of other curves was explained in Fig.~\ref{fig:loss cases}.
}
\label{fig:reduced field lossmaps}
\end{figure*}

This analysis shows that different loss channels can be observed in the loss map and this information can be used to deduce how different ITER components affect fast ion loss mechanics.
For example, similar analysis performed in Ref.~\cite{sarkimaki2018mechanics}, showed that the shift in divertor loads, caused by the plasma response to ECCs, was because plasma response suppressed stochastic-field-line transport while introducing the perturbed-banana loss channel.
On the other hand, the loss-map analysis can be used to increase confidence in the results of the orbit-following simulations.
Consider, for example, that loss-maps showed TBMs decrease ripple-induced transport while causing significant stochastic-field-line transport.
Both of these observations would run counter to what is known about the magnetic field structure (in~\ref{app:magnetic}) which shows that TBMs increase ripple magnitude but do not introduce field stochasticity.
This conflict would severely question the validity of the orbit-following simulation.

Finally, it is worth pointing out that the loss maps make a quantitative connection between different transport mechanisms and power losses.
In other words, they can be used to assess how much power is lost via different transport mechanisms.
This is achieved by plotting fraction of total power lost from a given bin, instead of fraction of particles lost, and integrating over the area of the loss channel.

\subsection{Reduced-field scenarios}

ITER will begin its operation in the pre-fusion phase with reduced field and current.
According to the ITER research plan\cite{iterresearch}, there are two scenarios, half-field ($I_p=7.5$ MA, $B_\phi=2.65$ T) and third-field ($I_p=5$ MA, $B_\phi=1.8$ T), which aim to achieve H-mode with helium plasma.
Both NBI and ICRF heating are required, but there is a limited understanding of fast ion confinement in these scenarios.
Here especially the ripple is of concern: 
The FIs over-compensate the TF ripple since they are optimized for the full-field operation and remain fully saturated.
Furthermore, TBMs are present as well.
As a consequence, the (over-compensated) ripple magnitude at the OMP separatrix is 1\% in the half-field scenario and 1.4\% in the third-field scenario (see~\ref{app:magnetic} for further details).
For reference, the ripple is 0.5\% in the full-field scenario.

While NBI ion confinement has been studied and found to be well-confined\cite{Kurki_Suonio_2009,Shinohara_2009,kurkibeam}, the assessment of ICRH ion confinement is severely lacking~\cite{schneider2018modelling}.
This is mainly because there is a lack of tools that couple ICRH generation to orbit-following.
However, we can perform loss-map analysis without knowing the details of ICRH ion distribution in the hopes that it provides some insight for the ICRH confinement.

We make the following assumptions based on the available data.
We assume the magnetic field and the equilibrium to be similar to that of the full-field.
The reduced field magnetic background, $\tilde{\mathbf{B}}$, can then be obtained via scaling
\begin{equation}
\tilde{\mathbf{B}} = \kappa\mathbf{B}_\mathrm{rip} + (\mathbf{B}_\mathrm{tot} - \mathbf{B}_\mathrm{rip}),
\end{equation}
where $\mathbf{B}_\mathrm{tot}$ refers to total full-field (including FIs and TBMs), $\mathbf{B}_\mathrm{rip}$ to the full-field with unmitigated ripple and no TBMs, and $\kappa$ is the scaling factor.
Temperature is assumed to be 10 keV and density $0.5n_G$ where $n_G$ is the Greenwald density.

As for the test particle population, we assume hydrogen ions with 1 MeV of energy.
We do not weight the markers but, instead, we populate the $(\rho',\xi')$ space uniformly to search the extent of the loss channels.

The loss-maps generated from the slowing-down simulations are shown in Fig.~\ref{fig:reduced field lossmaps}.
We have included also corresponding full-field scenario (with helium plasma and $n=0.5n_G$, $T=10$ keV) for comparisons sake.
First-orbit loss channel extends deeper as the field strength decreases and particle orbit-widths grow correspondingly larger.
However, it is the ripple-induced losses that become dominant extending as far inside as $\rho'=0.6$ in the third-field case.
But ICRH ions cannot be expected to be deeply trapped, most are ions whose banana tip is near the resonant surface.
Assuming the resonant surface is on axis, we can mark the corresponding particles on $(\rho',\xi')$-space, as was done in the loss maps in Fig.~\ref{fig:reduced field lossmaps}.
We note that these particles do not reside inside the actual loss-channel, i.e., inside the black 90\% contour, except at the very edge in the third-field scenario.

Since most of the ICRH power is expected to be deposited to ions near the core, we do not expect that the ripple causes collisionless losses of ICRH ions that would lead to unacceptable wall loads.
However, this can only be verified once the ICRH birth distribution is known.

\section{Prospects of using loss maps}
\label{sec:prospects}

Loss-maps can be used for other purposes as well than just for the post-simulation analysis.
They can used, e.g., to predict losses without carrying out orbit-following simulations, or to optimize marker generation further.
These other purposes are reviewed here.

\subsection{Connecting losses directly to magnetic field structure}

\begin{figure*}[!t]
\centering
\begin{overpic}[width=0.55\textwidth]{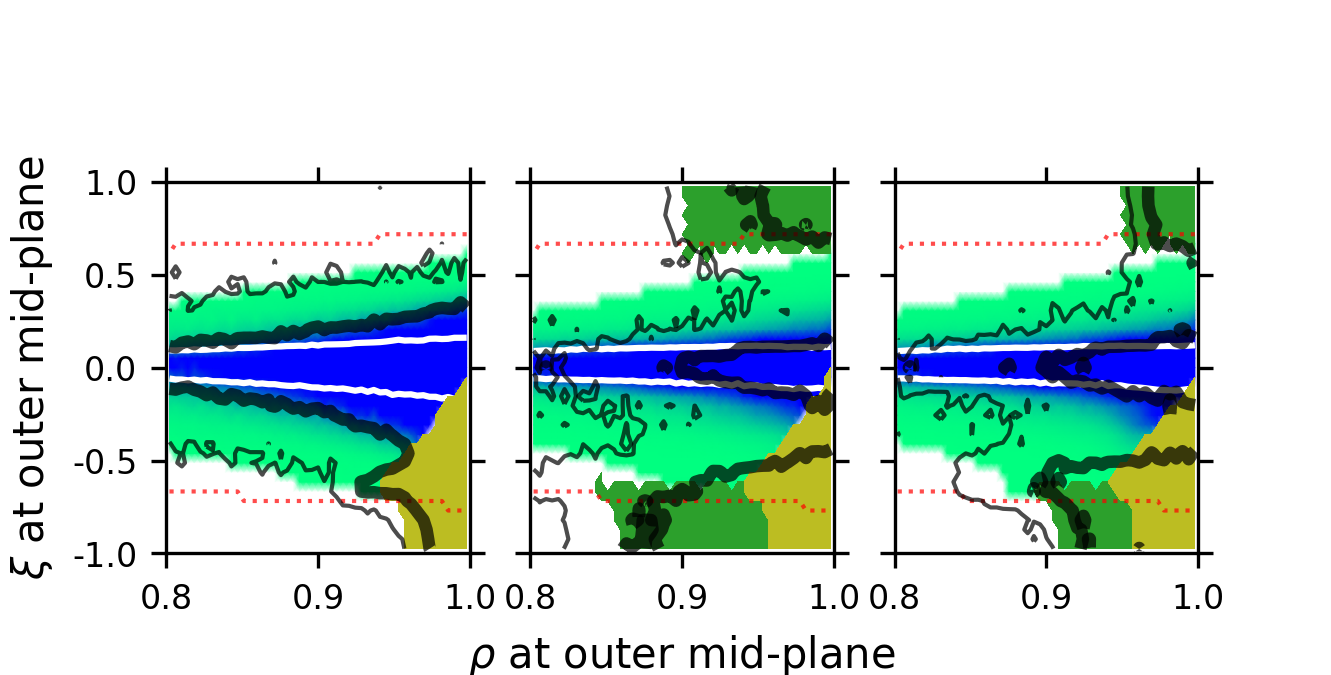}
\put(14,34){a)}
\put(41,34){b)}
\put(69,34){c)}
\end{overpic}
\caption{
Comparison between losses evaluated with orbit-following method and loss regimes deduced from magnetic field structure for cases (a) TF (b) +ECC, and (c) +PR.
Thick and thin black lines are the 10\% and 90\% loss contours from orbit-following calculations.
The ripple-trapping region (white contour), first-orbit-loss region (olive), stochastic-field-line transport region (green), and stochastic ripple-region (light green and blue) were evaluated from the magnetic field data.
The stochastic-ripple region shows the value of the diffusion coefficient: the coefficient is small ($D<0.1$ m$^2$/s) at the green area and large ($D>10$ m$^2$/s) at the blue region.
}
\label{fig:heuristic loss maps}
\end{figure*}

Of the cases explored in Fig.~\ref{fig:loss cases}, the magnetic field ripple is strongest in TF case, followed by +TBM, and weakest in +FI.
The loss maps showed that the ripple-induced losses followed the same pattern. 
One can even say how much power is lost due to the ripple in each case, but is that number as large, or small, as one would expect?

This is a common issue faced by modellers.
Since with ITER we cannot yet rely on the experiments, one way to verify the results would be to repeat them with multiple orbit-following codes.
However, here we show how we can make an estimate by contructing a loss map based on the magnetic field structure alone.
This can then compared to the one generated from simulation results to see, e.g., if the ripple-induced losses are as expected.

The construction is done by "projecting" the loss channels from $(R,z,\xi)$-space to the $(\rho',\xi')$-space.
First-orbit losses are projected by populating the $(R,z,\xi)$-space with alpha particles, choosing the ones that are outside the separatrix, and mapping those to the $(\rho',\xi')$-space.
Stochastic-field-line losses are projected in a same manner, by populating the $(R,z,\xi)$-space and mapping the chosen particles to the $(\rho',\xi')$-space, except that this time particles that are inside the ergodic field line region are chosen.
These regions can be identified with the help of the field-line Poincar\'e plots.
Furthermore, only particles that are passing are chosen as only those are affected by the stochastic-field-line transport.

The ripple-well, Eq.~\eqref{eq:ripple-well parameter}, and the critical stochastic-ripple, Eq.~\eqref{eq:critical ripple}, regions can both be plotted on a $Rz$-plane (see~\ref{app:magnetic} for illustrations).
By populating these regions with particles whose pitch is zero, i.e. their turning point is there, we can project them to the $(\rho',\xi')$-space.
Furthermore, we also project the value of the stochastic-ripple diffusion coefficient, Eq.~\eqref{eq:stochastic field diffusion}, via similar procedure.

We make these projections for the TF, +ECC and +PR cases whose loss maps were already presented in Fig.~\ref{fig:loss cases}.
These cases cover all transport mechanisms while containing variation in ripple strength and field stochasticity, making then ideal to illustrate the magnetic-field-based loss maps.
These loss maps are shown in the figure~\ref{fig:heuristic loss maps}.
The projections, indicated by the colors and the white contour, are shown along with the loss fraction (black contours) which was obtained from the orbit-following calculations.
Showing these in the same plot makes comparing the two convenient.

The TF case in Fig.~\ref{fig:heuristic loss maps}~(a) has no stochastic-field-line loss channel, so there we can compare the first-orbit projetion to the simulated losses.
Projection of the first-orbit losses (olive) is quite accurate if one compares the colored region with the black contour.
However, recalling the axisymmetric case in Fig.~\ref{fig:loss cases} (a), the first-orbit loss channel does not extent to narrow bananas ($\xi'<-0.5$) like the projection implies.
This is because the projection does not take into account that there is a gap between the wall and the separatrix, and the narrow bananas do not reach the wall.

The region where the stochastic-ripple transport criterion is met (light green) is almost equal in all cases.
One would be tempted to state that this projection can be used to predict losses as there is a good match in (a) between the green region and the 10\% loss contour.
Unfortunately, there is little agreement in (b) and (c) and, as such, we resist the temptation.
On the other hand, the region where the diffusion coefficient is large (blue) matches the 90\% loss contour in all cases.
It is difficult to say whether the ripple-well projection (white contours) can be used to project losses since it is enclosed by the blue region in all cases.

There seems to be a discrepancy in (b) and (c), where the region where the stochastic-ripple diffusion coefficient is significant (blue) extents to $\rho'< 0.9$, yet there are little losses there.
This is because particles do not travel vertically in $(\rho',\,\xi')$-space, not even when collisions are disabled and no scattering in $\xi'$ occurs.
When marker's $\rho'$ coordinate increases, $|\xi'|$ increases as well since the magnetic field (at the OMP) becomes weaker and magnetic moment is conserved.
Therefore, even though a particle born further inside the plasma has $\xi'$ close to zero and, thus, experiences ripple-induced transport, it escapes the loss channel if the channel is narrow as it is in (b) and (c).

The projection of the stochastic-field line region (dark green) falls between the 10\% and 90\% loss-fraction contours in Figs.~\ref{fig:heuristic loss maps} (b) and (c) where the mechanism is present.
One should note that in (c) there is a narrow region near $\xi'=-0.6$ where the loss fraction is over 90\% but the region is not covered in olive, blue, or dark green.
These losses are due to the perturbed banana transport, which were not projected since no similar criterion for it has been established yet as for other transport mechanisms.

We take the +PR case, Fig.~\ref{fig:heuristic loss maps} (c), as an example, and estimate the lost power based on the magnetic field structure and compare it to the orbit-following result.
We can estimate the losses by assuming that all particles born inside the olive, blue, and dark green regions are lost immediately.
Since the magnetic-field based estimate does not account for the collisional scattering, we only count those losses from the orbit-following simulation that happen in a collisionless time-scale, i.e., within 10 ms.
The magnetic-field based estimate gives 1.24 MW of losses while the orbit-following calculation gave 1.18 MW of lost power.

\subsection{Connecting birth location to wall loads}

\begin{figure*}[t]
\centering
\begin{overpic}[width=0.55\textwidth]{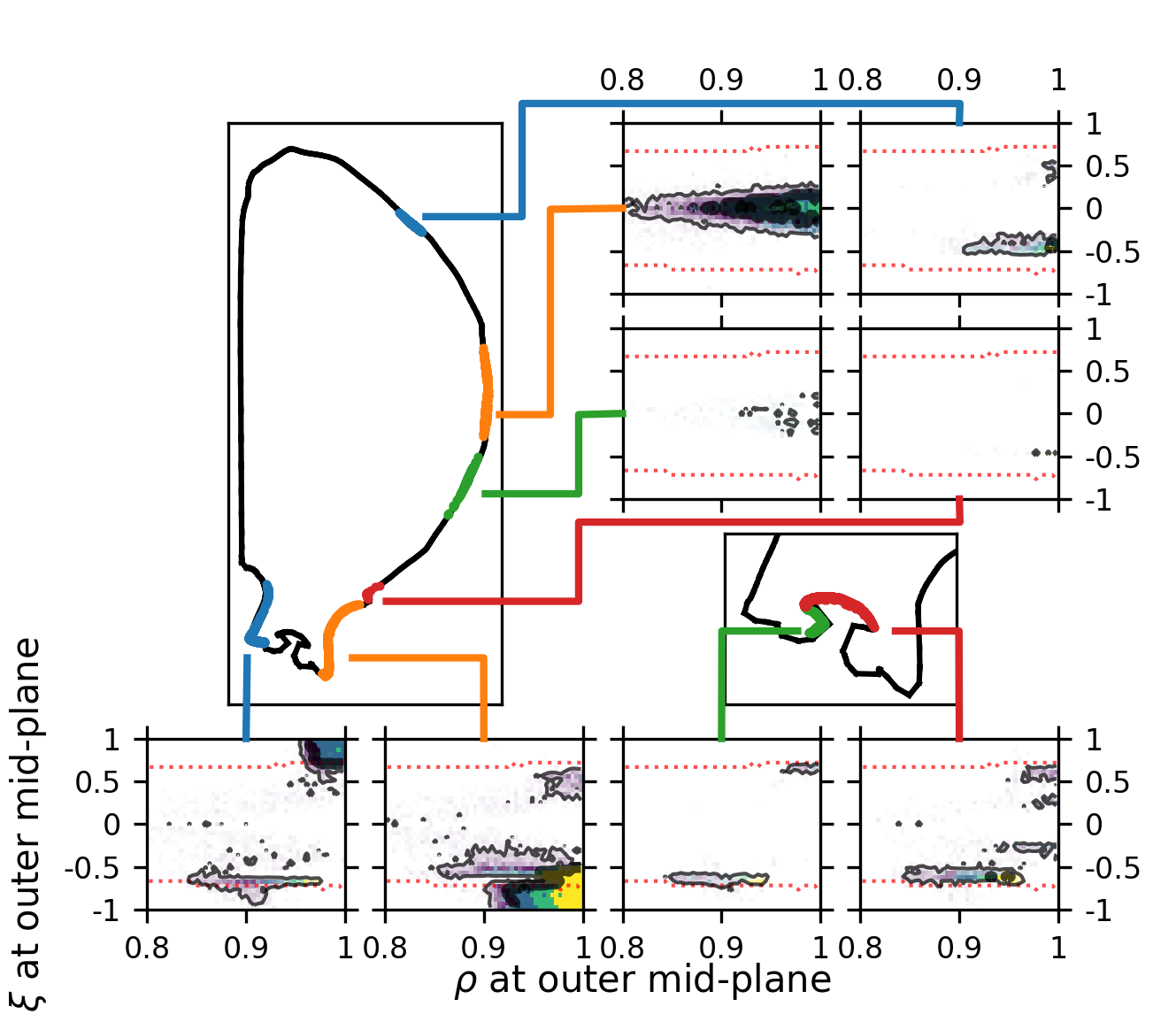}
\put(18,78){\scriptsize{i) Wall projection}}
\put(62,27.5){\scriptsize{ii)Divertor}}
\put(54,73){\colorbox{white}{\scriptsize{a)}}}
\put(74,73){\colorbox{white}{\scriptsize{b)}}}
\put(54,55){\colorbox{white}{\scriptsize{c)}}}
\put(74,55){\colorbox{white}{\scriptsize{d)}}}
\put(13,20){\colorbox{white}{\scriptsize{e)}}}
\put(33,20){\colorbox{white}{\scriptsize{f)}}}
\put(54,20){\colorbox{white}{\scriptsize{g)}}}
\put(74,20){\colorbox{white}{\scriptsize{h)}}}
\end{overpic}
\caption{
Connection between particles' birth locations in $(\rho',\xi')$-space and their final location on the wall or the divertor.
The wall projection is shown in (i) and the close-up of the divertor in (ii).
The four plots on the upper-left corner shows the loss maps for particles that have hit the specific region on the wall: 
(a) OMP losses,
(b) Upper wall losses,
(c) Below OMP losses,
(d) Bottom wall losses.
The divertor losses: (e) Inner leg losses, (f) Outer leg losses, (g) Under-the-dome losses, (h) Dome losses.}
\label{fig:wall loads}
\end{figure*}
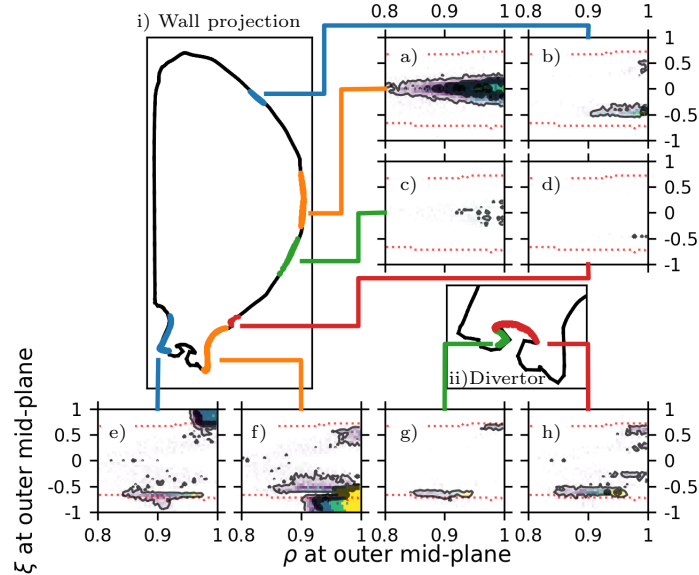

The converge study in Ref.~\cite{kurkisuonio2016effect} showed that $1\times10^6$ markers or more could be required to obtain accurate results when assessing peak power loads.
To compare, only $1\times10^4$ to $1\times10^5$ markers are required to obtain satisfying accuracy in total power losses or slowing-down distribution.
A similar issue plagues generation of synthetic FILD (Fast Ion Loss Detector) signal, because the detector is small and receives only small fraction of the fast ion population.
Here we show how a better converge can be achieved with fewer markers.
The loss maps can be utilized for this purpose by weighting the marker distribution in $(\rho',\xi')$-space so that more markers are born on loss channels (of interest).
These regions can be identified either by carrying out an exploratory simulation with few markers or by using the magnetic field based projection we described earlier.
Furthermore, if only markers at specific region in $(\rho',\xi')$ end up at the wall element that is of interest, marker distribution can be localised further increasing the signal-to noise ratio.

Figure~\ref{fig:wall loads} shows, using the +PR case as an example, how there is a connection between particle orbit topology, loss mechanism, and where on the wall it is lost.
We will discuss the wall losses, (a)~--~(d), first.
We can immediately notice that no passing particles are lost to the wall as they all end up to the divertor.
Most ripple-induced losses are lost to OMP (a), but some, probably those that were ripple-trapped, are lost under the OMP (c).
Some wider bananas are lost to the top (b) and some, although much fewer, are lost to the bottom region (d).

On the divertor, all co-passing particles are lost to the inner divertor leg (e) and counter-passing to the outer leg (f).
Some marginally trapped particles born with a negative pitch are lost to the inner leg (e) as well.
These are particles whose orbit on the high-field side would make an excursion below the X-point but, since they cannot cross there, they follow the field lines to the divertor inner leg.
The same thing happens for particles who, on the low-field side, try to cross the X-point below.
One can clearly observe that there is a "slit" in (f) near the trapped-passing region, and that the slit is left by the marginally trapped particles which end up to the inner leg (e).
The top edge of the hole divides trapped particles to those whose reflection point is at high-field side and to those who have it at low-field side.
There are also some particles in (f) with initially positive pitch that are lost to the inner leg via this process.

Finally, trapped particles contribute to the under-the-dome (g) and dome (h) losses.
Particles can reach the under-the-dome region, if they are reflected (second time) before they hit the inner divertor leg.
The particles lost on the dome are marginally trapped particles or wide banana particles.
Note that some of the wide banana particles were lost to (b) and so they do not contribute to the divertor losses.

One should note that this analysis is ITER specific, and the exact locations of the wall loads wary between machines and different operating scenarios.

\begin{figure}[t]
\centering
\begin{overpic}[width=0.45\textwidth]{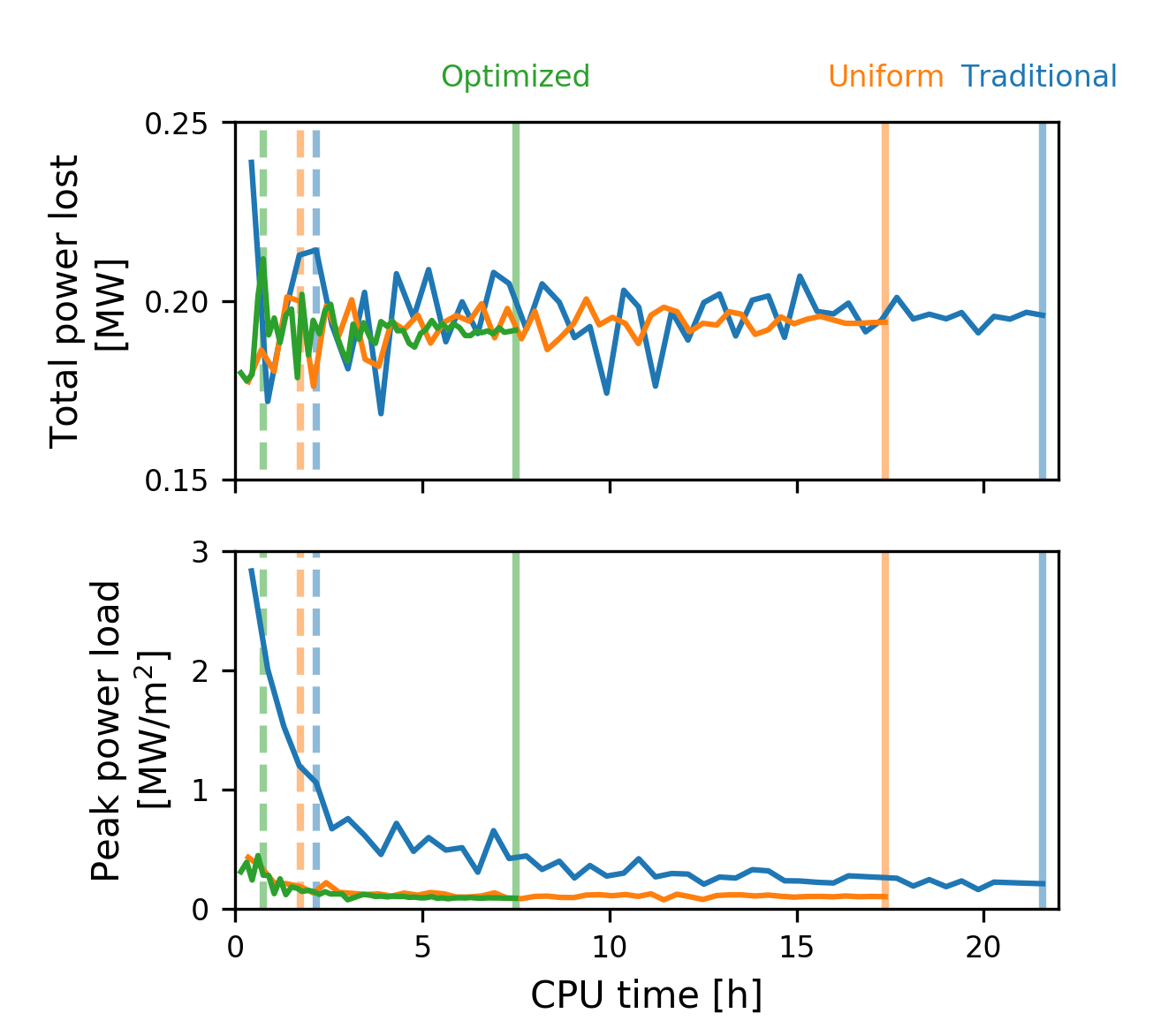}
\put(29,72){a)}
\put(29,35){b)}
\end{overpic}
\caption{
Convergence of (a) total power lost and (b) peak power load as a function of orbit-following calculation simulation time with different marker initialization procedures.
The solid vertical line indicates simulation has used $10^5$ markers while the dashed line is for $10^4$.
The curves were generated by sampling varying amounts of markers from a $10^5$ marker simulation.
}
\label{fig:convergence}
\end{figure}

We demonstrate what can be achieved with a better optimization by initializing markers in three different ways.
In the "traditional" way, markers are drawn directly from the particle distribution (Fig.~\ref{fig:initialization} (a)), i.e. they all have same weights, accepting those that are at the edge ($\rho>0.8$) for the simulation.
In the "uniform" way, markers are initialized uniformly in $(\rho',\xi')$, as we have done before when constructing the loss-maps, and accepting only the ones with $\rho'>0.8$.
Finally, the "optimized" way of initialization is performed using the loss map.
The markers are not initialized uniformly in $(\rho',\xi')$ but, instead, only on regions where there are losses.

Figure~\ref{fig:convergence} shows the convergence of total losses and peak power load on the wall when alpha particle losses were studied in the "+TBM" case (Fig.\ref{fig:loss cases} (d)).
Same number of markers were used in each case but, instead of plotting the convergence as a function of marker number, we show CPU time instead to reflect the fact that it takes less time to simulate markers that are lost.
While there is no dramatic difference in the convergence of the total losses, there is one in the convergence of the peak load estimate.
In fact, the traditional way have not saturated with $1\times10^5$ markers, as was expected, whereas "uniform" and "optimized" are.
Convergence of "uniform" is slightly better than "optimized" but one should keep in mind that the latter initialization requires that the loss-channels are located first.

\subsection{Role of collisions}

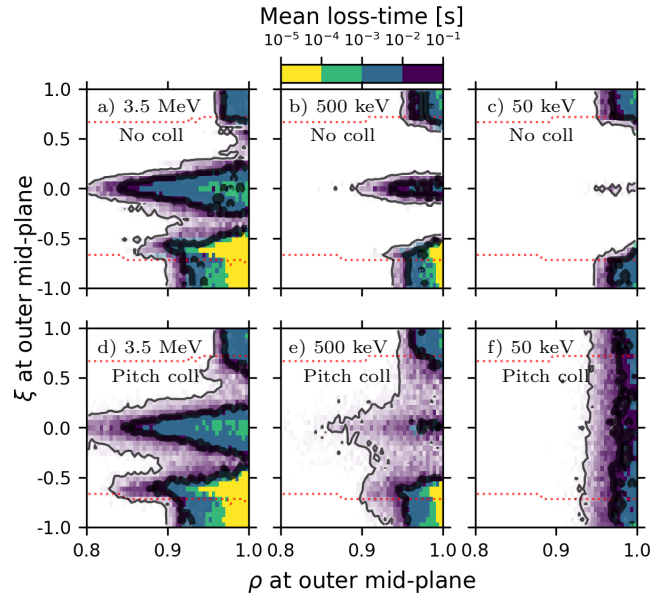
\begin{figure*}[t]
\centering
\begin{overpic}[width=0.55\textwidth]{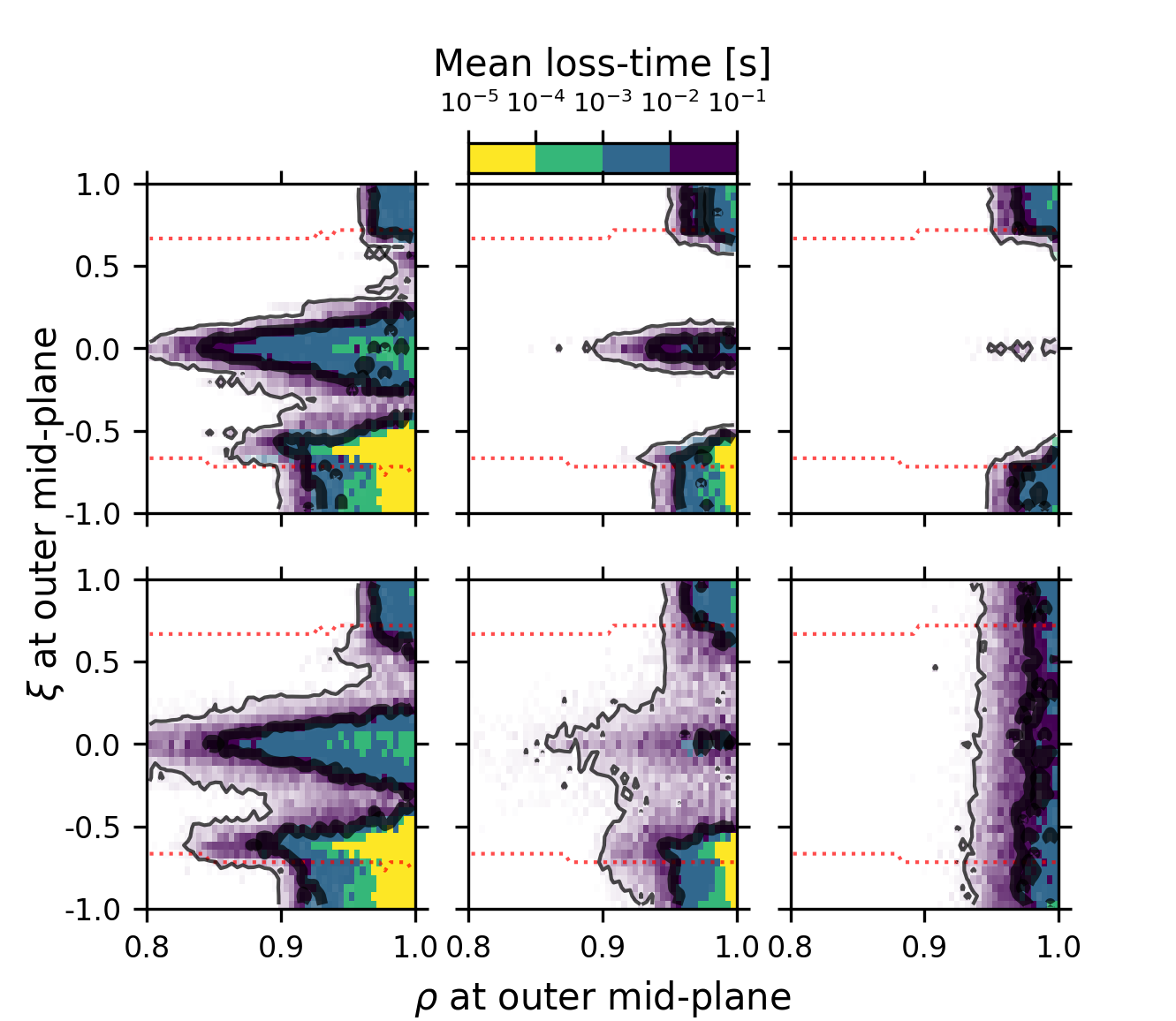}
\put(14,68){\scriptsize{a) 3.5 MeV}}
\put(17,64){\scriptsize{No coll}}
\put(41,68){\scriptsize{b) 500 keV}}
\put(44,64){\scriptsize{No coll}}
\put(69,68){\scriptsize{c) 50 keV}}
\put(72,64){\scriptsize{No coll}}
\put(14,34){\scriptsize{d) 3.5 MeV}}
\put(16,30){\scriptsize{Pitch coll}}
\put(41,34){\scriptsize{e) 500 keV}}
\put(43,30){\scriptsize{Pitch coll}}
\put(69,34){\scriptsize{f) 50 keV}}
\put(71,30){\scriptsize{Pitch coll}}
\end{overpic}
\caption{
Loss maps for alpha particles of different initial energies at +PR case.
Simulations were done twice: once without any collisions and once with just pitch collisions, i.e., with energy collisions disabled.
}
\label{fig:cases with collisions}
\end{figure*}

The loss maps have been used to illustrate loss channels due to collisionless transport mechanisms in slowing-down simulations which, naturally, involve collisions.
It is there interesting to explore what role, if any, collisions have on transport.
An alpha particle should experience little scattering initially since it is born above critical energy and, thus, it only experiences slowing-down due to collisions with electrons.
Therefore, collisional scattering should only become relevant around the critical energy when ion-ion collisions begin to dominate.
At thermal energies, scattering is so prelevant that particles may scatter in and out of the loss channels within few orbit periods, and the transport should be mostly neoclassical.

To verify this, the case +PR was simulated with three different alpha particle energies: birth energy 3.5 MeV, at 500 keV which is near the critical energy, and at thermal energy 10 keV.
Simulations were run for 0.1 s, and with two settings: one where collisions were disabled, and one where pitch scattering was enabled but energy remained constant.

Results are collected in Fig.~\ref{fig:cases with collisions} and they verify our expectations: at 3.5 MeV (a and d) collisions only slightly increase losses, becoming more important at 500 keV (b and e), and finally dominating at 10 keV (c and f) where losses no longer depend on pitch and the transport is neoclassical.
As a matter of fact, comparing 3.5 MeV collisionless case to the corresponding slowing-down run (Fig.~\ref{fig:loss cases} (f)) shows that the collisionless approximation can predict losses quite well.

\section{Fast ion transport as an advection-diffusion process}
\label{sec:advection diffusion}

Our motivation to investigate whether fast ion transport can be modelled as an advection-diffusion process comes from Ref.~ where this was shown to be true for runaway electrons.
However, runaway electrons are strongly passing so they are affected only by the stochastic-field-line transport whereas, for fast ions, all collisionless transport mechanisms are present.
Modelling transport as an advection-diffusion process serves two purposes.
First, this enables orbit-averaged codes to include transport due to 3D magnetic field by incorporating the transport coefficients.
Second, advection-diffusion model can be used to supplement orbit-following calculations to reduce overall simulation time, as we demonstrate later.

We begin by assuming that the transport can be modelled as one-dimensional process in $(\rho';\mu,E)$-space.
The magnetic moment and energy are treated as parameters for there is no transport in $\mu$ and $E$ if we neglect the collisions.
This simplification is justified at least for alpha particles since we showed, via orbit-following modelling, that collisionless loss maps are close to those obtained from the slowing-down simulations.
For comparisons with an orbit-following result, we can choose a population with a given energy, and scan the $(\rho',\mu)$-space with multiple simulations with different $\mu$ values.
Finally, the result is transformed to $(\rho',\xi')$-space.

Advection-diffusion processes are governed by the Fokker-Planck equation which, for the particle distribution function, $f=f(\rho',t;\mu,E)$, reads as
\begin{equation}
\frac{\partial f}{\partial t} = -\frac{\partial}{\partial \rho'}(K f) + \frac{\partial^2}{\partial\rho^{'2}}(Df),
\end{equation}
where $K(\rho';\mu,E)$ and $D(\rho';\mu,E)$ are the advection and the diffusion coefficient, respectively.
At the separatrix, $\rho'=1$, we set an absorbing boundary condition, $f(\rho')=0$.
The reflecting boundary condition can be placed at the core, $\rho'=0$, or at $\rho'$ value below which there is believed to be no significant transport.

This is a fairly trivial model which can be solved for a given initial distribution using e.g. Monte Carlo or Crank-Nicolson method.
But first the transport coefficients need to be evaluated with an orbit-following simulation which is less trivial.
This simulation is much shorter than the one used to estimate fast ion losses, since markers are only simulated for around tens of ploidal orbits~---~only long enough that the transport becomes apparent.

The coefficients are initially evaluated separately for each marker.
Then the $(\rho',\mu)$ domain is divided into bins and in each bin the markers' weighted average is used to represent $K(\rho';\mu)$ and $D(\rho';\mu)$.
The coefficient calculation is different for markers that are lost during this short simulation. 
For those markers the coefficients are evaluated from a (weighted) loss time distribution instead.
If a bin contains both lost and confined markers, the transport coefficients are evaluated separately for these and mean value, weighted with a fraction particles lost from a bin, is used in the end.

For confined markers, the coefficients are evaluated using similar means as in Ref.~\cite{Boozer_1981}.
Assuming that the transport is locally uniform, initially delta-peaked distribution evolves as
\begin{equation}
f(\rho',t) = \frac{1}{\sqrt{4\pi Dt}}\exp\left( -\frac{(\rho'-\rho'_0-Kt)^2}{4Dt} \right),
\end{equation}
and the coefficients can be evaluated as
\begin{align}
K &= \frac{\mathbb{E}[\rho'_i]-\rho'_0}{t},\\
D &= \frac{\mathbb{V}\mathrm{ar}[\rho'_i]}{2t},
\end{align}
where $\rho'_i$ are recorded data points in time $t$ for a single marker.
In practice there are few considerations.
First, continuous mapping of marker coordinates to $\rho'$ is too expensive to perform run-time and, therefore we avoid it by recording the marker position at outer mid-plane where $\rho=\rho'$.
For trapped particles, this means the sign of pitch must be checked and verified to be same in all crossings before the recording is made.
Secondly, we do not use the $\rho'_i$ values directly but take an average of $N$ subsequent crossings
\begin{equation}
\rho'_j = \frac{1}{N}\sum_i^N \rho'_i,
\end{equation}
and use those values instead to evaluate $K$ and $D$.
We do so to reduce noise that arise from the toroidally bent field lines.
Consider, for example, a passing particle in a field with just TF coil ripple.
There would be variation in $\rho'_i$ in subsequent OMP crossings and, consequently, $\mathbb{V}\mathrm{ar}[\rho'_i]$ would not be zero, and neither the diffusion coefficient, even though no actual transport takes place.
Choosing $N$ and $t$ is critical for success, and they can be deduced with some trial and error by evaluating coefficients with different values and carrying out comparisons between the advection-diffusion model and orbit-following simulation.

For lost markers, the coefficients are evaluated using the first-passage time.
The first passage time is the time in which a random walker, governed by the Langevin equation, first passes $\rho'_1$ when its initial location is $\rho'_0$.
Assuming $K>0$ and $\rho'_1 > \rho'_0$, the first-passage time is distributed as an \emph{Inverse-Gaussian}:
\begin{equation}
T(t) = \sqrt{\frac{c_2}{2\pi t^3}}\exp\left( -\frac{c_2(t-c_1)^2}{ 2c_1^2t } \right),
\end{equation}
where $c_1=\Delta\rho'/K$, $c_2=2(\Delta\rho')^2/D$, and $\Delta \rho' = \rho'_1-\rho'_0$.
The transport coefficients may now be evaluated as
\begin{align}
K &= \frac{\Delta\rho'}{\mathbb{E}[t_k]},\\
D &= \frac{2K^3 \mathbb{V}\mathrm{ar}[t_k]}{\Delta\rho'},
\end{align}
where $t_k$ is the time a marker $k$ was lost.

\subsection{Benchmark to orbit-following simulation}

\begin{figure}[b]
\centering
\begin{overpic}[width=0.3\textwidth]{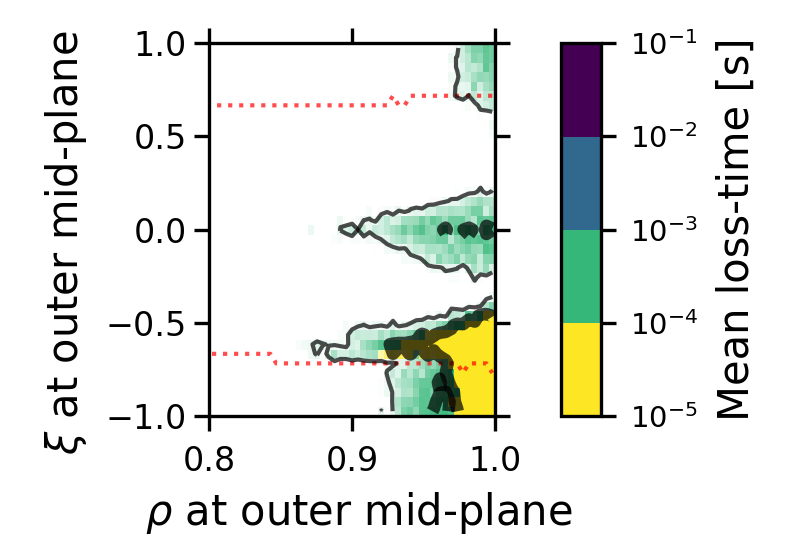}
\end{overpic}
\caption{
Loss map for the 1 ms simulation that was used to evaluate the transport coefficients.
The meaning of the different curves is the same as in Fig.~\ref{fig:loss cases}.
}
\label{fig:1ms simulation}
\end{figure}

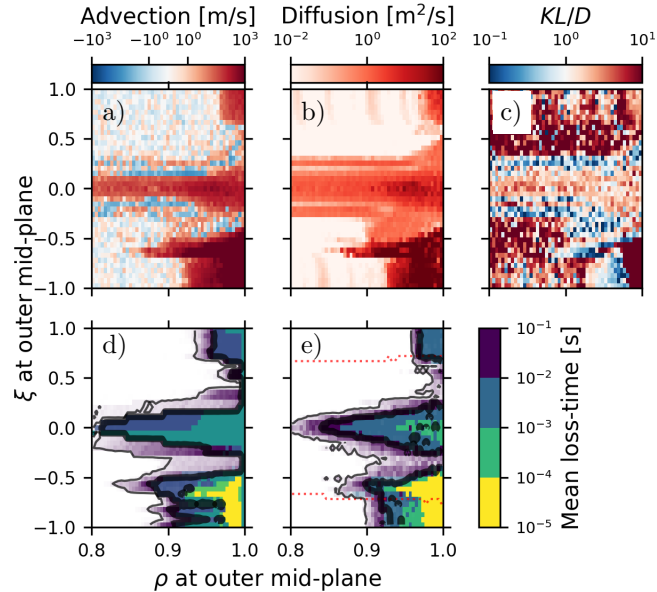
\begin{figure*}[t]
\centering
\begin{overpic}[width=0.55\textwidth]{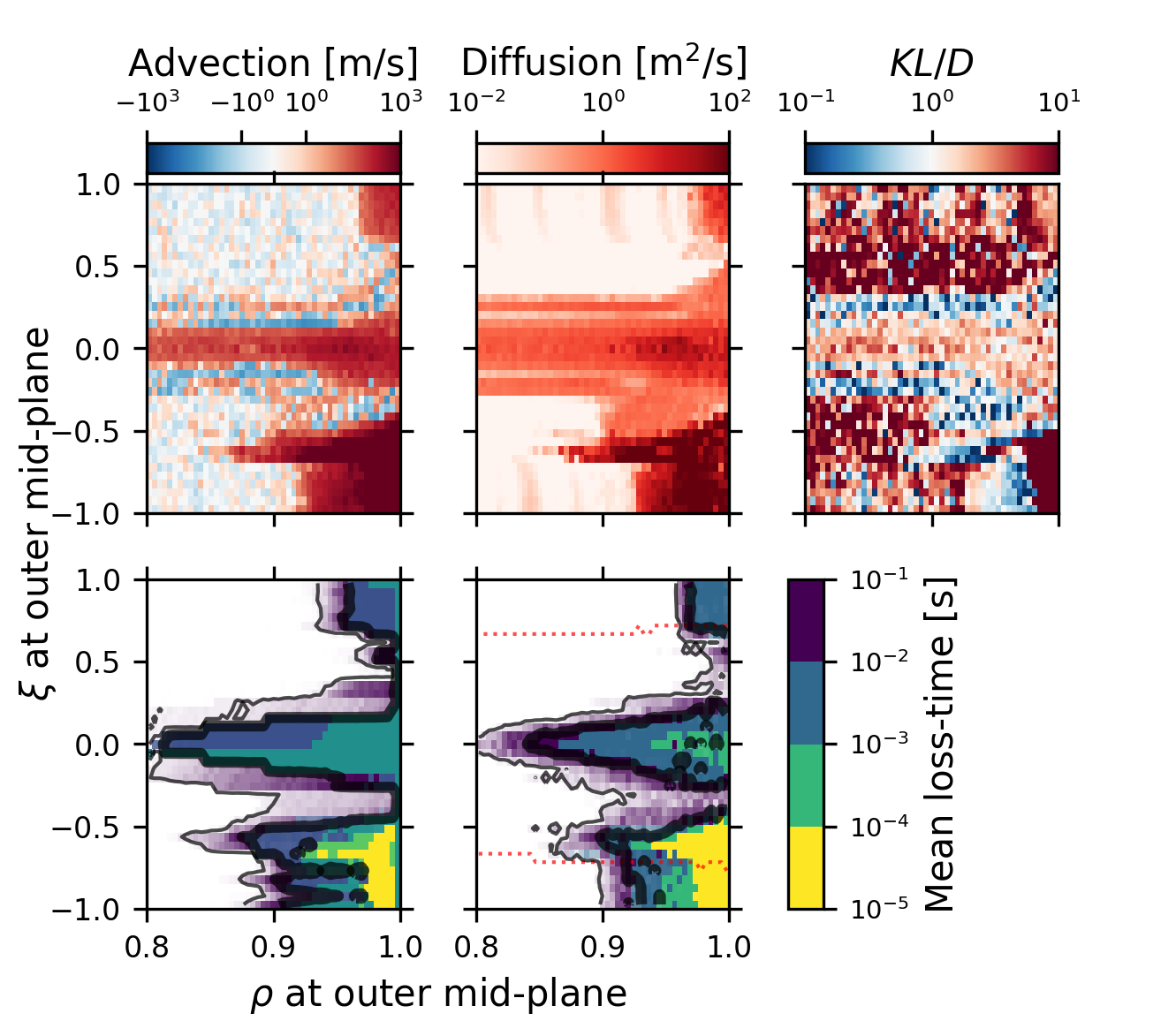}
\put(14,67){a)}
\put(42,67){b)}
\put(69,67){\colorbox{white}{c)}}
\put(14,34){d)}
\put(42,34){e)}
\end{overpic}
\caption{
Transport coefficients and model benchmark using +PR case as a testbed.
(a) Advection and (b) diffusion coefficients from an orbit-following simulation.
(c) P\'eclet number $P=KL/D$ which indicates whether transport is dominated by advection, $P>1$ (red regions), or diffusion, $P<1$ (blue).
For characteristic length scale, $L$, we have chosen $L=0.1$ m.
Loss map generated with (d) advection-diffusion model and with (e) orbit-following calculation.
}
\label{fig:transport model}
\end{figure*}

For benchmark purposes, the transport coefficients are evaluated for alpha particles in the +PR case because all transport mechanisms are present there, and since it is closest to what the actual magnetic field in ITER will be.
Orbit-following simulation is carried for $t=1$ ms, with collisions disabled, and averaging $N=10$ subsequent data points.
The loss map for this 1 ms simulation is shown in Fig.~\ref{fig:1ms simulation}, from which we can see that only the first-orbit loss channel has fully developed.
However, the transport coefficients provided by this simulation allow us to predict what the losses are going to be on a longer time scale, as we will soon show.

The evaluated transport coefficients are shown in Fig.~\ref{fig:transport model}, together with the result of the advection-diffusion model run for 100 ms.
The result is shown side-by-side with the corresponding orbit-following result, i.e., the one in Fig.~\ref{fig:loss cases}~(f).
We will discuss this comparison, Figs.~\ref{fig:transport model} (d) and (e), first before discussing what the coefficients imply.
Comparing the 90\% loss lines, the match seems to be quite accurate though the advection-diffusion model somewhat over-estimates the losses.
Evaluating the total lost power, we find that the advection-diffusion model predicts 2.46 MW of lost power whereas they are 1.79 MW in the collisionless orbit-following simulation.
What is remarkable is that losses can be predicted with such accuracy but using only 1/100th of the computational time required by the full orbit-following simulation.
Even though the model is not completely accurate and cannot predict wall loads, it is useful in cases in which a fast estimate is required or for scanning for interesting cases to be modelled with full orbit-following runs.
The accuracy increases if only those losses that occur within a collisionless time-scale ($t<1\times 10^{-2}$ s) are considered: orbit-following simulation gives 1.22 MW and advection-diffusion model 1.42 MW of lost power in this case.

As for the coefficients, the loss channels are clearly visible in both advection and diffusion plots.
Advection is strongest at the first-orbit loss region and, even though also diffusion is strong there, it is clearly an advection-dominated mechanism.
Both advection and diffusion are present for stochastic-field line transport, and it is interesting to note that there appear some regions with non-zero diffusion but where no losses originate.
Ripple-well region is advection dominated and it is surrounded by thin region where the advection is negative.
On stochastic-ripple regime advection has no clear form unlike the diffusion coefficient.
The perturbed banana transport appears as a spike in both advection and diffusion coefficients.
Coefficients are zero on regions we know no transport is present.

\subsection{Fast ion losses as a function ECC phasing}

\begin{figure*}[t]
\centering
\begin{overpic}[width=0.75\textwidth]{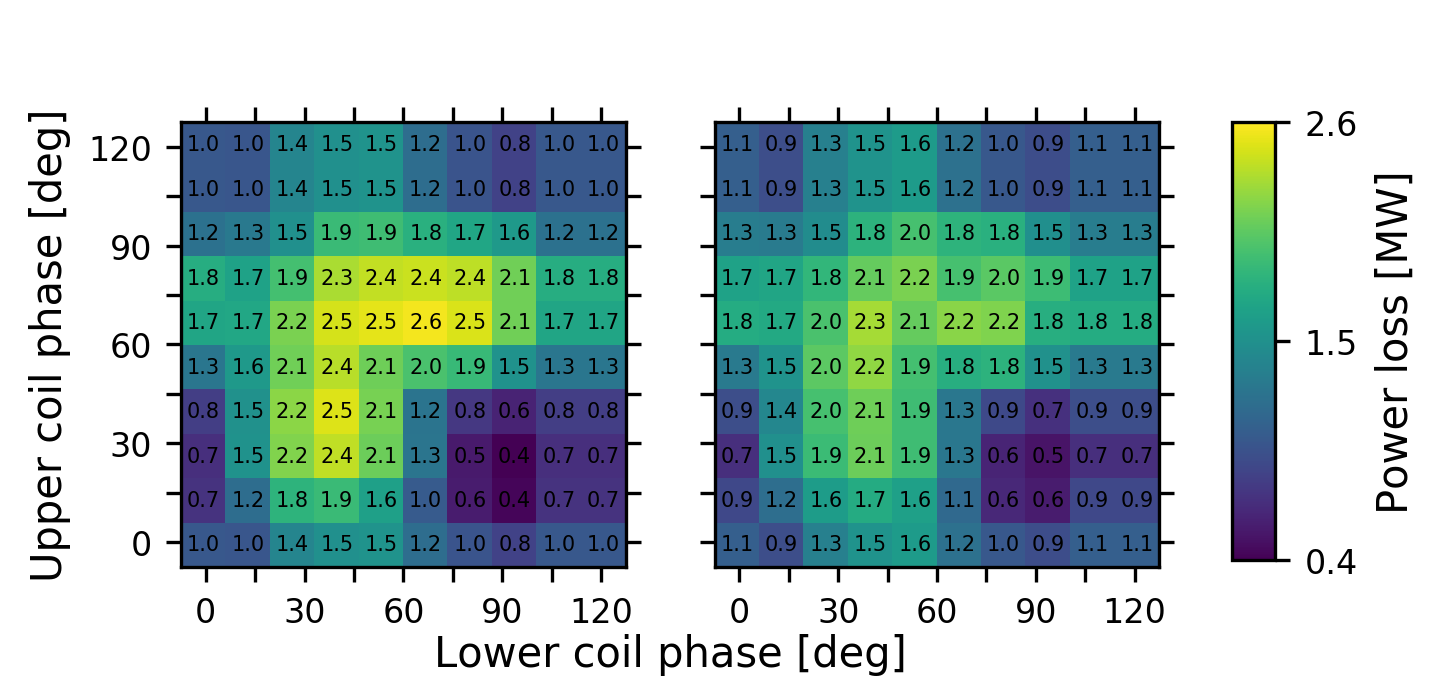}
\put(14,68){a) Orbit-following}
\put(42,68){b) Advection-diffusion model}
\end{overpic}
\caption{
Scan of alpha particle losses in ITER baseline scenario with ECC coil configuration $n=3$, $I=45$ kAt.
(a) Results orbit-following simulations where markers were simulated for the full slowing-down process.
(b) Results for advection-diffusion model where the transport was modelled for 100 ms, using the transport coefficients obtained by computing marker orbits for 1 ms.
The number in each bin indicates lost power in MW.
Calculations were done in a vacuum approximation.
}
\label{fig:rmpscan}
\end{figure*}

We demonstrate how the advection-diffusion model can benefit fast ion studies by using it to estimate fast ion losses due to ELM control coils (ECCs).
ITER hosts ECCs in three rows (upper, equatorial, and lower), each row having six coils, and each coil with a current
\begin{equation}
I_\mathrm{coil} = I \cos ( n [\phi_m - \phi_\mathrm{coil} ] ),
\end{equation}
where $I$ is the ECC current, $n$ the toroidal mode, $\phi_m$ the poloidal phase, and $\phi_\mathrm{coil}$ toroidal location of the coil.
Since $I$, $n$, and $\phi_m$ are not fixed, significant computational resources are needed to find where in the parameter space the fast ion confinement is acceptable.
That is if we were to assess fast ion losses using just the orbit-following tools.
With the advection-diffusion model, the scan can be performed with less need for resources, as we will demonstrate.

For this demonstration, we choose $I=45$ kAt and $n=3$ and scan alpha particle losses as a function of upper and lower coil phases.
The middle row phase is fixed to 0$^\circ$.
Perturbation due to ECCs is calculated with the code BioSaw.
Alpha particle losses are studied in the baseline scenario with FIs and TBMs present.
However, we do not include the plasma response here as it will be left for a dedicated study.

The scan is done for $8\times 8$ cases where the upper and lower coil phases are varied as $0^\circ$, $15^\circ$, $\dots$, $105^\circ$.
For each case, the orbit-following method is used to evaluate the transport coefficients.
The coefficients are evaluated in a $t=1$ ms simulation and the averaging is done for $N=10$ subsequent orbits.
The coefficients are substituted to the advection-diffusion model, which is then simulated for 100 ms to estimate the total losses.
The result is compared to a full slowing-down orbit-following simulation.

The comparison is shown in Fig.~\eqref{fig:rmpscan}.
The advection-diffusion model yields accurate losses for cases where losses are low, but underestimates the losses in cases where they are high.
However, the model accurately reproduces the overall shape, which can be described as consiting of two bands, $\phi_\mathrm{upper} = 90^\circ$ and $\phi_\mathrm{lower} = 60^\circ$, where losses are high.
The highest losses with both methods are estimated to be where these two bands cross.
In conclusion, the model can be used to find the interesting regions in parameter space and to give at least a rough estimate on losses.
Since the advection-diffusion model itself is cheap to simulate and all the cost is in evaluating the coefficients, scanning the losses takes 1/100th of the time required by the full orbit-following simulations.

\section{Conclusions}
\label{sec:conclusions}

We have shown how collisionless fast particle losses in 3D magnetic field can be directly linked to responsible transport mechanisms and magnetic field perturbations.
This is accomplished by carrying out an orbit-following simulation, and using the results to determine fraction of particles lost that are born in a given phase-space location.
Here we have used an intuitive set of phase-space coordinates, the pitch and radial position when a particle crosses the outer mid-plane, that uniquely defines the particle orbit topology (in 2D).
We have shown that, by evaluating fraction of particles lost in this phase-space, distinct loss channels appear.
This construction was dubbed as a loss map, and we demonstrated the many uses it has from improving signal-to-noise ration in orbit following simulation to better understanding of fast particle transport.

Orbit-following simulations are not needed to construct loss maps, since they can be constructed from the analytical estimates for the collisionless trasport mechanisms.
This allows one to estimate losses alternatively without carrying an orbit-following simulation, which provides confidence to the orbit-following results if there is an agreement.
We also demonstrated how loss-map analysis can help in understanding the underlaying mechanisms leading to fast particle losses, further providing confidence in ones results.

Another technique we demonstrated, was the procedure of treating fast ion transport as an advection-diffusion process.
We showed that, when the transport coefficients are evaluated with an orbit-following code, the fast ion transport can be solved with an 1D-model to a good accuracy.

We applied these techniques to study fast ion transport in ITER.
We showed how various transport mechanisms for alpha particles behave when the magnetic field is perturbed by TF coil ripple, ferritic inserts, test blanket modules, ELM control coils, and plasma response.
We made exploratory studies for ICRH confinement in the reduced field cases and, based on the results, we do not expect significant fast ion losses assuming most ICRH power is deposited to core ($\rho < 0.7$) ions.
With the advection-diffusion model we made a scan on alpha particle losses over ELM control coil phase configurations.
This scan was hundred times faster to perform than the full slowing-down simulations, which were done with an orbit-following code.
There was a decent agreement between the advection-diffusion model and orbit-following results, and we conclude that the former is suitable for scanning interesting regions in the ELM control coil parameter space.

We believe the techniques presented in this paper will prove useful for the fellow orbit-following modellers in reducing simulation time and increasing the confidence in ones results.

\ack
The work was funded by the Academy of Finland project No. 298126.
We acknowledge the CINECA award under the ISCRA initiative, for the availability of high performance computing resources and support.

\appendix

\section{Magnetic field structure in test cases}
\label{app:magnetic}

\begin{figure*}[t]
\centering
\begin{overpic}[width=0.55\textwidth]{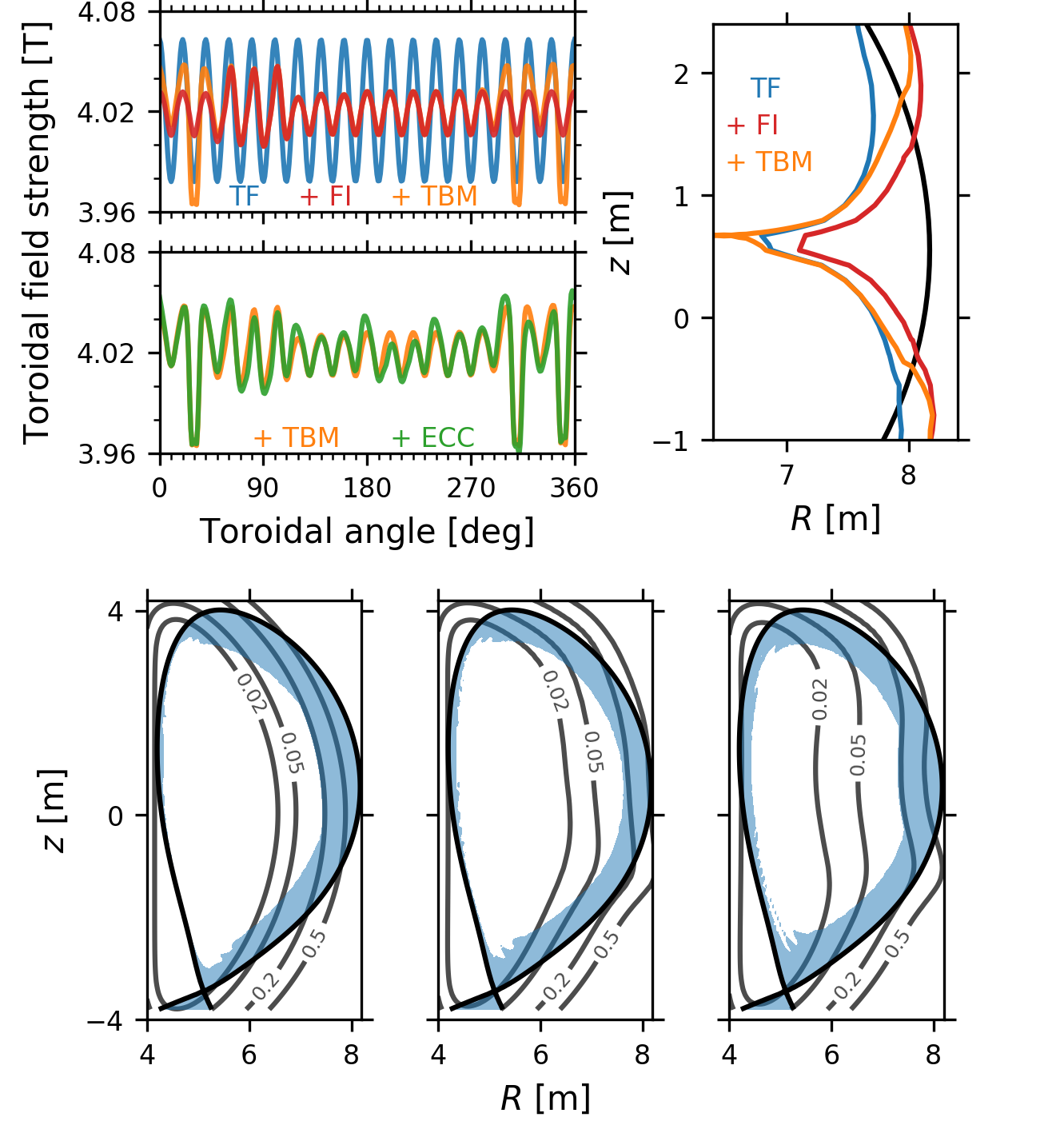}
\put(1,102){\footnotesize{a) Ripple at OMP separatrix}}
\put(58,102){\footnotesize{b) Ripple well}}
\put(1,49){\footnotesize{c) Ripple magnitude}}
\put(27,11){\scriptsize{TF}}
\put(51,11){\scriptsize{+FI}}
\put(75,11){\scriptsize{+TBM}}
\end{overpic}
\caption{
Illustration of the toroidal magnetic field ripple in different magnetic configurations. 
\textbf{a)} Variation of the toroidal field strength near the outer mid-plane separatrix, $(R=8.2,z=0.64)$, as a function of toroidal angle.
The top one plot shows the cases TF, +FI, +TBM, and the bottom one case +ECC, using +TBM as a reference.
\textbf{b)} Ripple wells. Colored lines limit the region where $a^*<1$ in each case.
\textbf{c)} Ripple magnitude (contours) and regions where stochastic-ripple transport criterion, $\delta > \delta_\mathrm{crit}$, is met for alpha particles with $\rho_g=5$ cm (blue).
Black curves show the separatrix location.
Even though the ripple is not periodic in +FI and +TBM cases, we have used $N=18$ in all cases when evaluating $\delta_\mathrm{crit}$.
}
\label{fig:ripple cases}
\end{figure*}

\begin{figure*}[t]
\centering
\begin{overpic}[width=0.55\textwidth]{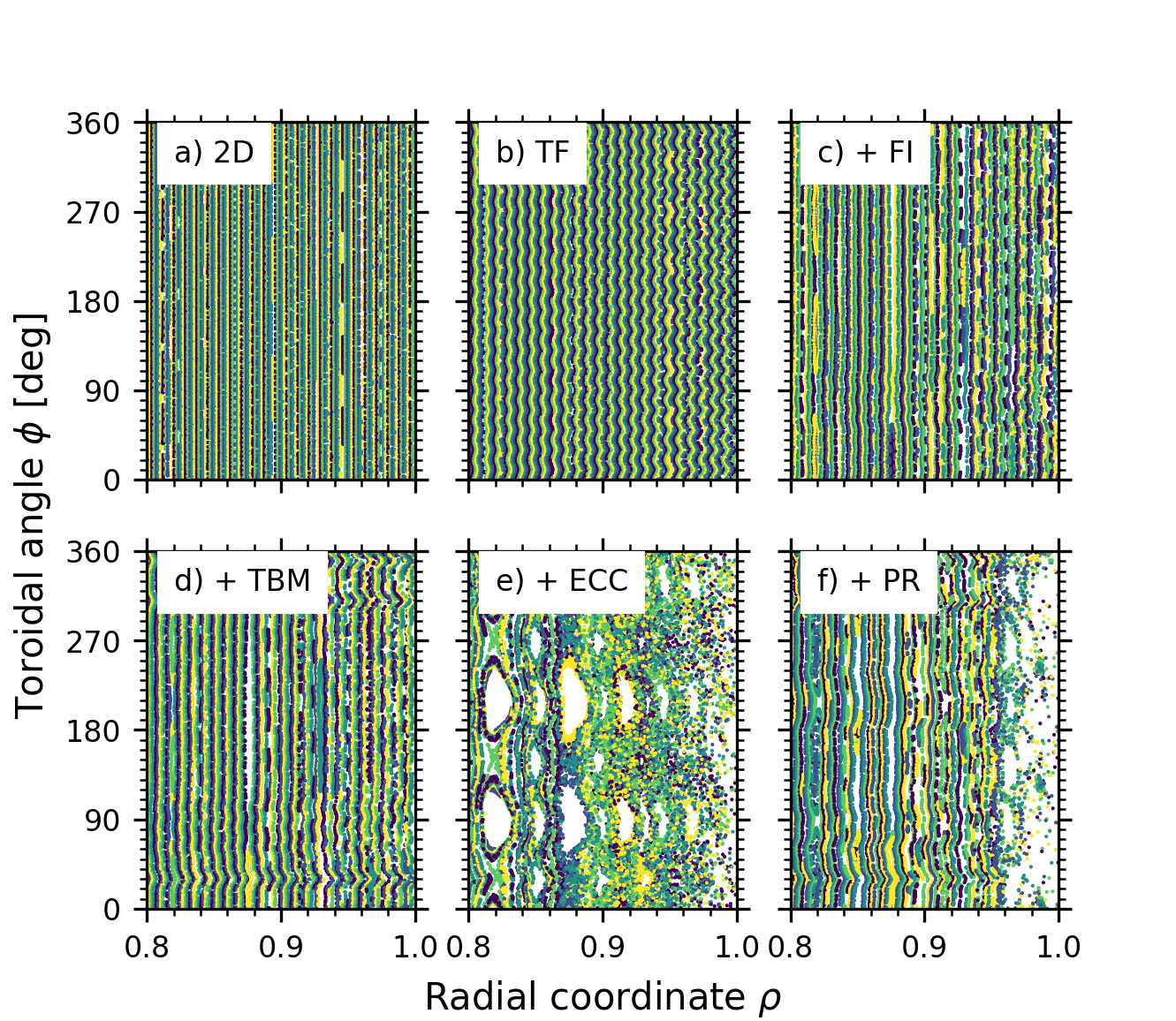}
\end{overpic}
\caption{
Magnetic field structure in different magnetic configurations illustrated with Poincar\'e plots.
A Poincar\'e plot is constructed by following a field-line marker and marking down the location each time the marker passes through a pre-defined plane.
Here several field lines were traced and the their $(\rho,\phi)$-coordinates were marked each time they passed OMP.
Colors are used to separate different field lines.
}
\label{fig:poincare cases}
\end{figure*}

\begin{figure*}[p]
\centering
\begin{overpic}[width=0.55\textwidth]{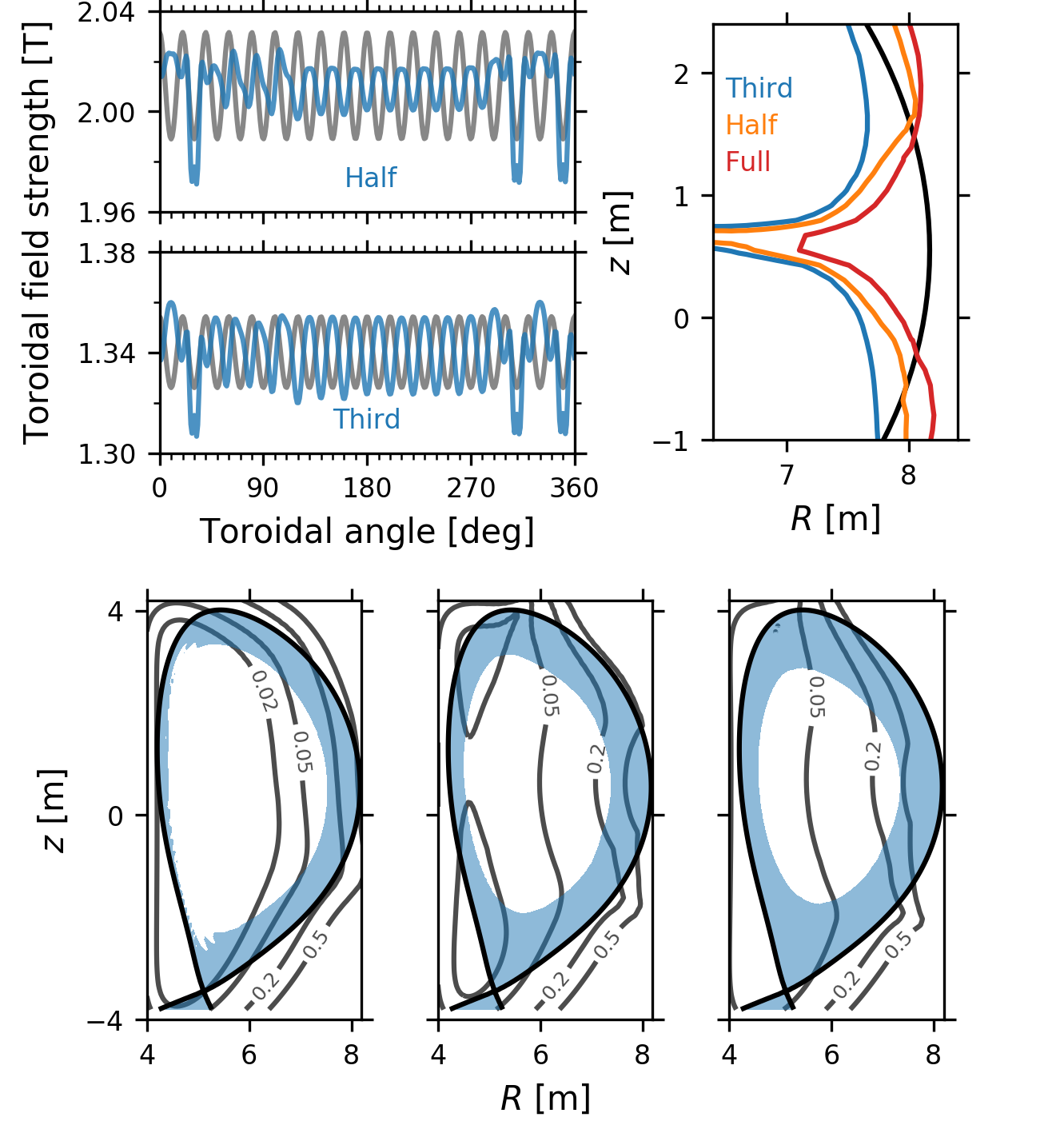}
\put(1,102){\footnotesize{a) Ripple at OMP separatrix}}
\put(58,102){\footnotesize{b) Ripple well}}
\put(1,49){\footnotesize{c) Ripple magnitude}}
\put(26,11){\scriptsize{Full}}
\put(52,11){\scriptsize{Half}}
\put(76.5,11){\scriptsize{Third}}
\end{overpic}
\caption{
Illustration of the toroidal magnetic field ripple in reduced field scenarios.
Plots have the same meaning as in Fig.~\ref{fig:ripple cases} except for the differences that we list here.
\textbf{a)} Variation of the toroidal field strength shows the ripple with FIs and TBMs in half-field (top) and third-field (bottom) scenarios.
The unmitigated ripple in both cases is also shown (grey curve).
\textbf{b)} Ripple wells.
\textbf{c)} Ripple magnitude and regions where stochastic-ripple transport criterion is met for (ICRH) hydrogen ions with $E=1$ MeV.
}
\label{fig:ripple reduced cases}
\end{figure*}

The loss-map analysis was demonstrated in section~\ref{sec:applying loss maps} for ITER baseline scenario with various magnetic field perturbations present.
The magnetic field structure in these cases is visualized in Figs.~\ref{fig:ripple cases} and~\ref{fig:poincare cases}.
Figure~\ref{fig:ripple cases}~(a) shows the general overview of the ripple.
ECCs and plasma response only slightly modify the ripple, which is why the latter is not even displayed here.
Ripple-well regions, according to ripple-well parameter Eq.~\eqref{eq:ripple-well parameter}, are plotted in Fig.~\ref{fig:ripple cases}~(b).
Ripple magnitude, Eq.~\eqref{eq:ripple definition}, and regions where ripple exceeds critical ripple value for stochastic-ripple transport, Eq.~\eqref{eq:stochastic ripple diffusion}, are on display in Fig.~\ref{fig:ripple cases}~(c).
Poincar\'e plots in Fig.~\ref{fig:poincare cases} are constructed by tracing field lines and marking down the locations where the field line cross OMP.
The ripple in reduced-field scenarios is depicted in Fig.~\ref{fig:ripple reduced cases}.

\section*{References}
\bibliographystyle{iopart-num}
\bibliography{lossmap}

\providecommand{\newblock}{}
\begin{thebibliography}{10}
\expandafter\ifx\csname url\endcsname\relax
  \def\url#1{{\tt #1}}\fi
\expandafter\ifx\csname urlprefix\endcsname\relax\def\urlprefix{URL }\fi
\providecommand{\eprint}[2][]{\url{#2}}

\bibitem{varje2016}
Varje J, Asunta O, Cavinato M, Gagliardi M, Hirvijoki E, Koskela T,
  Kurki-Suonio T, Liu Y, Parail V, Saibene G, Sipil\"a S, Snicker A,
  S\"arkim\"aki K and \"Ak\"aslompolo S 2016 {\em Nuclear Fusion\/} {\bf 56}
  046014 doi:
  \href{https://doi.org/10.1088/0029-5515/56/4/046014}{10.1088/0029-5515/56/4/046014}

\bibitem{varjeascot5}
Varje J, S\"arkim\"aki K, Kontula J, Ollus P, Kurki-Suonio T, Snicker A,
  Hirvijoki E and \"Ak\"aslompolo S In press {\em Journal\/}

\bibitem{akers2012gpgpu}
Akers R, Verwichte E, Martin T, Pinches S and Lake R 2012 {GPGPU Monte Carlo}
  calculation of gyro-phase resolved fast ion and n-state resolved neutral
  deuterium distributions {\em Proc. 39th EPS Conference on Plasma Physics\/}
  p~P5 \urlprefix\url{http://ocs.ciemat.es/epsicpp2012pap/pdf/P5.088.pdf}

\bibitem{akers2018high}
Akers R, Colling B, Hess J, Liu Y, Turner A, {\"A}k{\"a}slompolo S, Varje J,
  S{\"a}rkim{\"a}ki K, Pinches S and Singh M 2018 High fidelity simulations of
  fast ion power flux driven by 3d field perturbations on iter {\em Proc. 26th
  IAEA Fusion Energy Conference\/} pp TH/4--1
  \urlprefix\url{https://nucleus.iaea.org/sites/fusionportal/Shared%20Documents/FEC%202016/fec2016-preprints/preprint0489.pdf}

\bibitem{Kurki_Suonio_2009}
Kurki-Suonio T, Asunta O, Hellsten T, Hynönen V, Johnson T, Koskela T,
  Lönnroth J, Parail V, Roccella M, Saibene G, Salmi A and Sipilä S 2009 {\em
  Nuclear Fusion\/} {\bf 49} 095001 doi:
  \href{https://doi.org/10.1088/0029-5515/49/9/095001}{10.1088/0029-5515/49/9/095001}

\bibitem{Tani_2011}
Tani K, Shinohara K, Oikawa T, Tsutsui H, Miyamoto S, Kusama Y and Sugie T 2011
  {\em Nuclear Fusion\/} {\bf 52} 013012 doi:
  \href{https://doi.org/10.1088/0029-5515/52/1/013012}{10.1088/0029-5515/52/1/013012}

\bibitem{sarkimaki2018mechanics}
S\"arkim\"aki K, Varje J, B{\'{e}}coulet M, Liu Y and Kurki-Suonio T 2018 {\em
  Nuclear Fusion\/} {\bf 58} 076021 doi:
  \href{https://doi.org/10.1088/1741-4326/aac393}{10.1088/1741-4326/aac393}

\bibitem{White_1996}
White R~B, Goldston R~J, Redi M~H and Budny R~V 1996 {\em Physics of Plasmas\/}
  {\bf 3} 3043--3054 doi:
  \href{https://doi.org/10.1063/1.871641}{10.1063/1.871641}

\bibitem{Hsu_1992}
Hsu C~T and Sigmar D~J 1992 {\em Physics of Fluids B: Plasma Physics\/} {\bf 4}
  1492--1505 doi: \href{https://doi.org/10.1063/1.860060}{10.1063/1.860060}

\bibitem{Goldston_1981}
Goldston R~J, White R~B and Boozer A~H 1981 {\em Physical Review Letters\/}
  {\bf 47} 647--649 doi:
  \href{https://doi.org/10.1103/physrevlett.47.647}{10.1103/physrevlett.47.647}

\bibitem{Tobita_1992}
Tobita K, Tani K, Neyatani Y, van Blokland A~A~E, Miura S, Fujita T, Takeuchi
  H, Nishitani T, Matsuoka M and Takechi S 1992 {\em Physical Review Letters\/}
  {\bf 69} 3060--3063 doi:
  \href{https://doi.org/10.1103/physrevlett.69.3060}{10.1103/physrevlett.69.3060}

\bibitem{Rechester_1978}
Rechester A~B and Rosenbluth M~N 1978 {\em Physical Review Letters\/} {\bf 40}
  38--41 doi:
  \href{https://doi.org/10.1103/physrevlett.40.38}{10.1103/physrevlett.40.38}

\bibitem{akaslompolo2015iter}
\"Ak\"aslompolo S, Kurki-Suonio T, Asunta O, Cavinato M, Gagliardi M, Hirvijoki
  E, Saibene G, Sipil\"a S, Snicker A, S\"arkim\"aki K and Varje J 2015 {\em
  Nuclear Fusion\/} {\bf 55} 093010 doi:
  \href{https://doi.org/10.1088/0029-5515/55/9/093010}{10.1088/0029-5515/55/9/093010}

\bibitem{kurkisuonio2016protecting}
Kurki-Suonio T, S\"arkim\"aki K, \"Ak\"aslompolo S, Varje J, Liu Y, Sipil\"a S,
  Asunta O, Hirvijoki E, Snicker A, Ter\"av\"a J, Cavinato M, Gagliardi M,
  Parail V and Saibene G 2016 {\em Plasma Physics and Controlled Fusion\/} {\bf
  59} 014013 doi:
  \href{https://doi.org/10.1088/0741-3335/59/1/014013}{10.1088/0741-3335/59/1/014013}

\bibitem{kurkisuonio2016effect}
Kurki-Suonio T, \"Ak\"aslompolo S, S\"arkim\"aki K, Varje J, Asunta O, Cavinato
  M, Gagliardi M, Hirvijoki E, Parail V, Saibene G, Sipil\"a S and Snicker A
  2016 {\em Nuclear Fusion\/} {\bf 56} 112024 doi:
  \href{https://doi.org/10.1088/0029-5515/56/11/112024}{10.1088/0029-5515/56/11/112024}

\bibitem{Parail_2013}
Parail V, Albanese R, Ambrosino R, Artaud J~F, Besseghir K, Cavinato M,
  Corrigan G, Garcia J, Garzotti L, Gribov Y, Imbeaux F, Koechl F, Labate C,
  Lister J, Litaudon X, Loarte A, Maget P, Mattei M, McDonald D, Nardon E,
  Saibene G, Sartori R and Urban J 2013 {\em Nuclear Fusion\/} {\bf 53} 113002
  doi:
  \href{https://doi.org/10.1088/0029-5515/53/11/113002}{10.1088/0029-5515/53/11/113002}

\bibitem{akaslompolo2015calculating}
\"Ak\"aslompolo S, Asunta O, Bergmans T, Gagliardi M, Galabert J, Hirvijoki E,
  Kurki-Suonio T, Sipil\"a S, Snicker A and S\"arkim\"aki K 2015 {\em Fusion
  Engineering and Design\/} {\bf 98-99} 1039--1043 doi:
  \href{https://doi.org/10.1016/j.fusengdes.2015.05.038}{10.1016/j.fusengdes.2015.05.038}

\bibitem{liu2016modelling}
Liu Y, \"Ak\"aslompolo S, Cavinato M, Koechl F, Kurki-Suonio T, Li L, Parail V,
  Saibene G, S\"arkim\"aki K, Sipil\"a S and Varje J 2016 {\em Nuclear
  Fusion\/} {\bf 56} 066001 doi:
  \href{https://doi.org/10.1088/0029-5515/56/6/066001}{10.1088/0029-5515/56/6/066001}

\bibitem{Evans_2013}
Evans T, Orlov D, Wingen A, Wu W, Loarte A, Casper T, Schmitz O, Saibene G,
  Schaffer M and Daly E 2013 {\em Nuclear Fusion\/} {\bf 53} 093029 doi:
  \href{https://doi.org/10.1088/0029-5515/53/9/093029}{10.1088/0029-5515/53/9/093029}

\bibitem{iterresearch}
 2018 {ITER Research Plan within the Staged Approach (Level III - Provisional
  Version)} Tech. rep. ITER Organization
  \urlprefix\url{https://www.iter.org/technical-reports}

\bibitem{Shinohara_2009}
Shinohara K, Oikawa T, Urano H, Oyama N, Lonnroth J, Saibene G, Parail V and
  Kamada Y 2009 {\em Fusion Engineering and Design\/} {\bf 84} 24--32 doi:
  \href{https://doi.org/10.1016/j.fusengdes.2008.08.040}{10.1016/j.fusengdes.2008.08.040}

\bibitem{kurkibeam}
Kurki-Suonio T, S{\"a}rkim{\"a}ki K, Snicker S, Schneider M, Polevoi A and
  Mitteau R 2018 {Beam Ion Performance and Power Loads in the ITER Pre-Fusion
  Power Operating Scenarios (PFPO) with Reduced Field and Current} {\em {27th
  IAEA Fusion Energy Conference}\/}
  \urlprefix\url{https://conferences.iaea.org/indico/event/151/papers/5739/files/5154-Beam_loads_PFPO1.pdf}

\bibitem{schneider2018modelling}
Schneider M, POLEVOI A, Kim S, Loarte A, ARTAUD J, Beaumont B, Bilato R,
  Boilson D, Campbell D, Dumortier P {\em et~al.\/} 2018

\bibitem{Boozer_1981}
Boozer A~H 1981 {\em Physics of Fluids\/} {\bf 24} 851 doi:
  \href{https://doi.org/10.1063/1.863445}{10.1063/1.863445}

\end{thebibliography}

\end{document}